%% file: main.tex
\documentclass[conference,compsoc]{IEEEtran}
\IEEEoverridecommandlockouts

\usepackage{cite}
\usepackage{amsmath,amssymb,amsfonts}
\usepackage{algorithmic}
\usepackage{graphicx}
\usepackage{textcomp}
\usepackage{xcolor}
\usepackage{float}
\usepackage{caption}
\usepackage{setspace}
\usepackage{tabularx}
\usepackage[hyphens]{url}        
\usepackage{hyperref}             
\hypersetup{
    colorlinks=false,     
    hidelinks,            
    breaklinks=true       
}
\usepackage{etoolbox}
\appto\UrlBreaks{\do\-}           
\appto\UrlBreaks{\do\_}           
\appto\UrlBreaks{\do\/}           
\def\BibTeX{{\rm B\kern-.05em{\sc i\kern-.025em b}\kern-.08em
    T\kern-.1667em\lower.7ex\hbox{E}\kern-.125emX}}
\begin{document}

\title{Where Do LLM-based Systems Break? A System-Level Security Framework for Risk Assessment and Treatment}

\author{\IEEEauthorblockN{ Neha Nagaraja}
\IEEEauthorblockA{\textit{School of Informatics, Computing, and Cyber Systems} \\
\textit{ Northern Arizona University}\\
Flagstaff, USA \\
nn454@nau.edu}
\and
\IEEEauthorblockN{Hayretdin Bahsi\textsuperscript{1,2}}
\IEEEauthorblockA{\textsuperscript{1}\textit{School of Informatics, Computing, and Cyber Systems} \\
\textit{Northern Arizona University}, Flagstaff, USA \\
\textsuperscript{2}\textit{Department of Software Science} \\
\textit{Tallinn University of Technology}, Tallinn, Estonia \\
hayretdin.bahsi@nau.edu}
}

\maketitle

\input{abstract}


\input{sections/introduction}
\input{sections/relatedworks}
\input{sections/methods}
\input{sections/results}
\input{sections/discussion}

\input{sections/conclusion}

\begingroup
\footnotesize
\setstretch{0.8} 
\bibliographystyle{IEEEtran}
\bibliography{main}
\endgroup

\input{appendix}

\end{document}

%% file: abstract.tex
\textbf{Abstract - Large Language Models (LLMs) are increasingly integrated into safety-critical workflows, yet existing security analyses remain fragmented and often isolate model behavior from the broader system context. This work introduces a goal-driven risk assessment framework for LLM-powered systems that combines system modeling with Attack–Defense Trees (ADTrees) and Common Vulnerability Scoring System (CVSS)-based exploitability scoring to support structured, comparable analysis. We demonstrate the framework through a healthcare case study, modeling multi-step attack paths targeting intervention in medical procedures, leakage of electronic health record (EHR) data, and disruption of service availability. Our analysis indicates that threats spanning (i) conventional cyber, (ii) adversarial ML, and (iii) conversational attacks that manipulate prompts or context often consolidate into a small number of dominant paths and shared system choke points, enabling targeted defenses to yield meaningful reductions in path exploitability. By systematically comparing defense portfolios, we align these risks with established vulnerability management practices and provide a domain-agnostic workflow applicable to other LLM-enabled critical systems.}

%% file: sections/introduction.tex

\section{Introduction}

Large Language Models (LLMs) are increasingly deployed into safety-critical systems across domains, from healthcare to public infrastructure. Their ability to process natural language and orchestrate external tools makes them promising for automating complex workflows. Healthcare providers exemplify this shift, deploying LLM-powered assistants behind orchestration layers that coordinate external services (e.g., EHR backends, translators, and knowledge bases) to support tasks such as drafting notes, summarizing Electronic Health Records (EHRs), and assisting with clinical decision-making~\cite{openai2024gpt4, denecke2024}. While this promise is attractive, it also introduces new security and privacy risks: multi-step interactions among the LLM, external services, and healthcare infrastructure create complex attack surfaces~\cite{zhui2024} that are not well captured by traditional threat analyses.

Risk assessment is a core activity in the secure development life cycle (SDLC) ~\cite{MicrosoftSDL2025}: it translates a growing set of threats and weaknesses into prioritized engineering work by identifying which attack paths are feasible, which assets and goals they endanger, and which controls provide the paramount risk reduction under practical constraints. This need becomes more acute for LLM-enabled applications, mainly orchestrated or agent-based systems, because the attack surface extends beyond a single model or API call to include tool invocation, workflow logic, memory and state, and cross-component trust boundaries. Consequently, security analysis must account for how conventional platform attacks, adversarial ML attacks, and conversational attacks (e.g., prompt injections) interact and compound across an end-to-end system.

Most existing LLM security work focuses on model- or API-centric threats such as prompt injection~\cite{Rossi2024}, jailbreaks~\cite{Chu2024}, unsafe content generation, or training-data extraction, often studied in isolation and outside of concrete domains. At the same time, the security development life cycle community has mature practices around threat modeling and vulnerability scoring (e.g., STRIDE~\cite{Shostack20141}, CVSS~\cite{cvss}). Still, these are typically applied to conventional software components (web applications, databases, and networks) rather than to LLM-based systems that maintain cross-session memory and call out to tools. More broadly, many threat modeling and risk assessment efforts conducted at early stage enumerate individual threats but do not clearly demonstrate how those threats align into end-to-end attack paths that connect an entry point to a concrete security goal. Attack-path reasoning matters because defenses are often deployed at shared chokepoints (e.g., authentication boundaries, orchestration logic, and tool connectors), and path-based models make it explicit where a control breaks the chain. 

The challenge of risk assessment is amplified in LLM-enabled architectures, where three categories of threats must be consolidated within a single system view ~\cite{ICISSP2025, healthsec2026}: (i) conventional cyber threats that target the platform layer (e.g. MitM, credential theft), (ii) adversarial ML threats that exploit the deployed model, and (iii) conversational threats that induce harmful behavior through manipulated prompts or context (e.g., prompt injection). For instance, a white-box adversarial attack may require the attacker to compromise an infrastructure system component to learn the model parameters, or an indirect prompt-injection attack can be realized by compromising an information flow or third-party knowledge source. The literature focuses on individual adversarial or conversational threats. However, a consolidated threat view that captures the entire attack path is necessary to conduct a realistic risk assessment and to prioritize treatment in LLM-based systems.    

Attack graphs have long been used to represent attacker actions and paths in conventional systems \cite{lallie2020review}. These approaches rely on detailed deployment configurations (e.g., known vulnerabilities and network access control rules) that are often available only in later stages of development or after deployment. In contrast, early-stage analysis for complex ML-based architectures must remain usable under partial system knowledge, while still producing actionable attack paths and defensible prioritization. On the other side, attack trees provide more flexibility in terms of node descriptions in identifying the attack scenarios \cite{khalil2024threat}. However, no systematic approach has been proposed to identify and align tree nodes. The risk calculations conducted on the paths derived from the trees do not specify how to map the impact assessment to the goal of the attack path. Prioritization of risk treatment options has attracted less attention. The literature provides limited, system-wide risk assessment methods that unify heterogeneous threat classes for LLM-enabled systems, supporting defense selection and comparison.

This paper addresses this gap by adopting an attack-path-centric methodology for early-stage risk assessment: we treat the LLM as one component in a system architecture and explicitly model how heterogeneous threats compose into multi-step paths that impact mission goals. We focus on three security goals that are central to LLM deployments in healthcare ~\cite{healthsec2026}: (\(G_1\)) \emph{Intervening in medical procedures} (unsafe or manipulated clinical guidance), (\(G_2\)) \emph{Leakage of EHR data} (violations of patient confidentiality), and (\(G_3\)) \emph{Disruption of access or availability} (denial or degradation of timely access to records and decision support). For each goal, we construct attack–defense trees (ADTs) that capture realistic multi-step paths involving users, the web application, the orchestrator, the LLM, and external tools. We use CVSS-based exploitability scoring to estimate and compare the feasibility and likelihood of compound paths and identify where defenses most effectively break them. The tree nodes, except for the root, determine the likelihood of risk scenarios through a consolidated exploitability calculation for the corresponding attack paths, whereas the root node identifies the risk impact, as the security violation goal is specified in that node. This approach places the complete attack path as the center of risk calculation to estimate the impact with accordance to the goal of whole path. Our risk treatment approach also benefits from attack paths to provide a more systematic view of treatment options and their interdependencies. While our evaluation is instantiated as a healthcare case study, the methodological workflow is domain-agnostic. It applies to other LLM-enabled critical systems by substituting goals, assets, and system components.

Concretely, we organize our study around the following research questions:
\emph{RQ1:} How can attack–defense trees be used to model multi-step attacks that exploit interactions between LLM components and surrounding system infrastructure and tools at the early stage of system development, considering all threat categories, including conventional, adversarial, and conversational ones?
\emph{RQ2:} How can CVSS exploitability scores be mapped onto, and aggregated across, attack–defense tree structures to estimate compound risk from heterogeneous threats?
\emph{RQ3:} Under realistic defense budget constraints, which combinations of controls provide the most significant reduction in path exploitability per unit cost, and how can our framework guide these choices?

This paper makes three contributions:
(1) \emph{Goal-driven system modeling and attack-path construction.} We present a goal-driven modeling method for LLM-enabled critical systems that combines system-level data flow diagrams with attack–defense trees, capturing (i) conventional platform threats, (ii) adversarial ML threats, and (iii) conversational threats under goals (\(G_1\))–(\(G_3\))in a healthcare case study.
(2) \emph{Exploitability scoring for multi-step attack paths.} We show how to attach CVSS v3.1 exploitability vectors to ADT leaves (via representative CVEs and explicit LLM-specific assumptions where CVEs do not exist) and propagate them through OR/AND/SAND compositions to obtain path-level exploitability scores that reflect compound, multi-step attacks.
(3) \emph{Defense-portfolio comparison under cost constraints.} We introduce a risk-treatment workflow that models concrete security controls as transformations on CVSS base metrics, defines simple cost levels for those controls, and evaluates canonical hardening scenarios (precondition-first, guardrails-first, and combined) on a realistic healthcare deployment.

The novelty of this work is methodological: it operationalizes attack-path-centric risk assessment for LLM-enabled systems by unifying heterogeneous threat classes within a single system model, producing explicit paths that support defense placement and portfolio comparison under partial, early-stage knowledge. By connecting paths to CVSS scoring and a control-and-cost workflow, it bridges the gap between abstract AI security concerns and established vulnerability management practices. The framework provides a practical, domain-agnostic risk management approach for securing LLM deployments across critical infrastructures.


%% file: sections/relatedworks.tex
\section{Related works}

As LLMs become embedded in safety- and mission-critical systems; spanning healthcare, finance, critical infrastructure, and autonomous operations structured security risk assessment has struggled to keep pace with adoption. The security literature on LLM-integrated systems remains fragmented: despite rapid adoption, structured risk assessment is still limited, and prior work primarily emphasizes STRIDE- or stakeholder-based threat taxonomies (e.g., prompt injection, data leakage, hallucination) without connecting individual threats to multi-step attack scenarios or operational impact~\cite{Tete2024, Pankajakshan2024,hamid2024,clusmann2024}. In parallel, traditional IT risk frameworks and outage analyses capture important infrastructure realities, including risks in mission-critical networked systems and cascading failures driven by shared IT dependencies and market concentration. Still, they are often static and coarse-grained, omitting attacker behavior, attack paths, and cross-workflow propagation that are central to operational and patient-safety consequences~\cite{aijaz2023,yurcik2024}. Complementary lines of work in IoT/embedded threat modeling and trust-centered studies similarly provide valuable perspectives on cyber-physical exposure and patient–AI decision dynamics. Yet, they under-model AI-native behaviors, rapidly evolving software vulnerabilities, and system-level threat propagation across interconnected components~\cite{omotosho2019,vakhter2022,10538960,298114}. Meanwhile, structured attack-path reasoning has matured in other domains: attack-defense trees have been applied to ATM security and ML pipelines. They can support explainable risk reasoning even for non-experts~\cite{fraile2016,hoseini2024,schiele2025}, and NIST CSWP 35 ~cite{} further illustrates how system modeling, STRIDE enumeration, and attack trees can be applied to genomic sequencing. However, these established approaches primarily target conventional cyber threats in comparatively static infrastructure and do not fully capture the heterogeneous, behavior-driven threat landscape introduced by LLM-integrated systems, motivating goal-driven, system-level risk modeling adaptable to any domain where LLM orchestration governs critical workflows. 

The Common Vulnerability Scoring System (CVSS) provides a standardized framework for quantifying vulnerability severity and has become the authoritative metric in vulnerability management.  Although CVSS v3.1/v4.0 has been used for medical device threat modeling and semi-quantitative risk scoring, it has not yet been extended to system-level LLM orchestration where threats propagate across components~\cite{10538960}. Bahar and Wazan ~\cite{bahar2024} demonstrate that CVSS v3.1 exhibits insufficient metric variability when applied to individual LLM attacks in isolation. However, this limitation can be reframed at the system level. By consolidating individual threats into multi-step orchestration paths via attack-defense trees, we shift exploitability assessment from metric factors alone to path-level and cross-component interactions. This enables CVSS to provide meaningful differentiation where isolation-level metrics cannot, while exposing chokepoints where defenses break end-to-end attacks.

Collectively, this leaves a gap: a lack of structured, system-wide models that quantify how attackers can combine conventional cyber threats, adversarial ML techniques, and LLM-specific exploits along actionable paths to compromise critical operations. This work addresses that gap by proposing a goal-driven, attack-defense tree framework tailored to LLM-orchestrated healthcare systems, integrating calibrated CVSS metrics to expose dominant attack paths and critical defensive chokepoints.

%% file: sections/methods.tex
\section{Methodology}

In this section, we formalize our end-to-end risk assessment workflow for LLM-powered healthcare systems. This workflow integrates and advances two distinct strands of our prior research. First, our system modeling work ~\cite{ICISSP2025} produced the system model and a foundational catalog of component-level threats (via STRIDE). Second, our concurrent study \cite{healthsec2026} used high-level attack trees to prioritize risk scenarios using a likelihood × impact scheme, providing a high-level prioritization view. In contrast, the present study focuses on exploitability and risk treatment: it operationalizes path-level feasibility scoring and enables explicit comparison of defense portfolios under cost constraints.

Specifically, we (i) construct goal-driven Attack–Defense Trees (ADTs) for three security objectives; (ii) aggregate exploitability across multi-step attack paths using CVSS v3.1 scoring and combine these with goal-specific impact profiles; and (iii) define canonical defense scenarios and model their effects as transformations of CVSS metrics at selected leaves. This workflow enables systematic comparison of risk reduction and implementation cost across heterogeneous controls.

\subsection{System Modeling}

We analyze a representative LLM-powered healthcare assistant that answers patient and clinician queries, surfaces EHR context, and coordinates external tools. Why this matters for risk: the model anchors our Attack–Defense Tree (ADT) analysis in concrete interfaces, data flows, and trust boundaries, both constraining the attack surface we score and locating where defenses can actually be deployed. 

\begin{figure}
    \centering
    \includegraphics[width=1\linewidth]{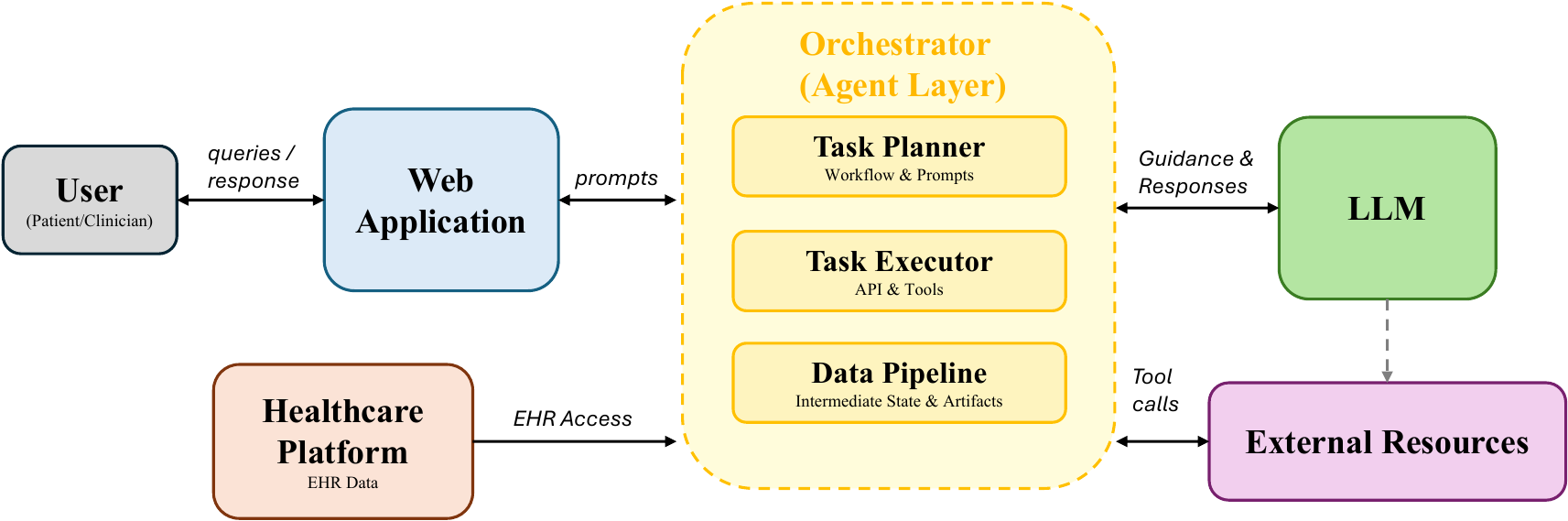}
    \caption{System Architecture of the LLM-based Healthcare Assistant}
    \label{fig:sysarch}
\end{figure}

Core Components ~\cite{ICISSP2025}. Figure~\ref{fig:sysarch} 
illustrates the system architecture. A \textbf{Web Application} mediates user sessions; a \textbf{Healthcare Platform} supplies governed EHR data; an \textbf{Orchestrator} (agent layer) sequences tools and routes calls via a \textit{Task Planner} (workflow blueprint + prompt scaffolds), \textit{Task Executor} (API/tool invocation), and a \textit{Data Pipeline} (intermediate state, artifacts); \textbf{External Resources} (e.g., translators, clinical KBs, analytics) provide enrichment; the \textbf{LLM} performs the reasoning pass and returns guidance. The LLM may be hosted or third-party, and may be fine-tuned or adapted; this broadens the attack surface beyond pure inference, which is reflected in our ADT construction and risk treatment.

\subsection{Threat Modeling}

In earlier work ~\cite{ICISSP2025} on the same LLM-based healthcare assistant, we performed a complete STRIDE-per-element ~\cite{ICISSP2025, Shostack20141} analysis over the system-level DFD. Based on this analysis, we constructed an attack taxonomy by combining entries from MITRE ATLAS ~\cite{ATLAS2025} with the OWASP Top 10 for LLMs ~\cite{OWASP2024} and conventional infrastructure threats. The resulting threats clustered into three families: (i) conventional cyber threats that target the platform layer (e.g., MitM, credential theft), (ii) adversarial ML threats that exploit the deployed model, and (iii) conversational threats that induce harmful LLM behavior through manipulated prompts or context (e.g., prompt injection). We then mapped this taxonomy onto eight security boundaries across the components and their interactions, treating every cross-boundary flow as a potential attack vector, and cataloging concrete threats at each interface. In this paper, we reuse that taxonomy, along with its boundary assignments, as the starting point for constructing our attack–defense trees and quantifying risk. 

\subsection{Systematic Risk Assessment}

Our risk assessment is goal-driven: instead of identifying risks of individual threats in isolation, we model how attackers achieve three high-level security goals within attack paths ~\cite{healthsec2026}, G1: intervening in medical procedures, G2: leakage of EHR data, and G3: disruption of access or availability, and how defenses alter these threat paths.
We structure this analysis in three steps. First, we build Attack–Defense Trees (ADTs) for each goal, organizing threats into explicit preconditions, execution steps, and resulting impacts, providing a compositional view of how attacks unfold across system components [RQ1](see subsection \ref{attack-defense}). Second, we apply CVSS- based ~\cite{cvss} exploitability scores to ADT and aggregate them using logical connectors, yielding quantitative exploitability scores for complete attack paths [RQ2] (see subsection\ref{cvss-method}). Finally, we evaluate risk treatment by attaching concrete controls and recomputing path exploitability, enabling comparison of mitigation strategies and trade-offs between residual risk and implementation cost [RQ3] (see subsection \ref{treat-method}).

\subsection{Attack–Defense Tree Modelling}
\label{attack-defense}

We model risk using goal-driven Attack–Defense Trees (ADTs) rather than isolated threats. For each security goal ~\cite{healthsec2026}, \textit{G1: Intervening in Medical Procedures}, \textit{G2: Leakage of EHR Data}, and \textit{G3: Disruption of Access or Availability}, we construct a separate tree whose root encodes the attacker’s high-level objective. Internal nodes describe how this objective can be realized through combinations of lower-level threats, and attached defense nodes represent concrete controls that disrupt or harden those paths. Each ADT is instantiated directly from the attack taxonomy developed in our prior work, so every node corresponds to a previously identified threat or control.

We explicitly decompose each attack path into three semantic layers ~\cite{healthsec2026}: \textit{Preconditions} (\textit{P}) capture what must already have gone wrong before a specific exploit becomes usable, i.e., how the attacker first gains leverage over the system. For example, without compromising the orchestration workflow, an attacker cannot misroute tasks or poison shared memory. \textit{Execution} (\textit{V}) represents the active attack given those preconditions, i.e, observable attack patterns such as submitting a malicious query, injecting unauthorized tasks, or hijacking a user session. \textit{Final impact} maps a successful execution step back to one of our three goals, such as unsafe clinical guidance, EHR confidentiality breach, or denial of access to patient records.

This precondition–execution–impact decomposition is central to our risk assessment. It lets us distinguish (i) controls that make it harder to reach the attack surface (e.g., stronger authentication, session isolation), (ii) controls that make it harder to exploit once the surface is reached (e.g., prompt guardrails, orchestration consistency checks), and (iii) controls that limit the resulting damage (e.g., narrowing the scope of accessible data).

We use three logical connectors to express how attacks compose. \texttt{OR} nodes encode alternative strategies where any child suffices (e.g., user hijacking, \texttt{OR} MitM, \texttt{OR} malicious user as ways to compromise the prompt channel). \texttt{AND} nodes encode joint requirements where all children must hold, but order is irrelevant (e.g., a poisoning attack requiring both LLM trained on sensitive EHR data \texttt{AND} model access exposed). Sequential AND (\texttt{SAND}) encodes ordered dependencies: some conditions must hold before others. At the path level, we model every attack as a \texttt{SAND} ~\cite{Konsta2024AttackTreesGraphs} of preconditions then execution (\textit{P} \texttt{SAND} \textit{V}), reflecting that exploit attempts only matter once the necessary footholds exist.

Defenses are attached as counter-nodes to specific preconditions or execution steps. A defense node denotes a family of controls (e.g., MFA, mTLS, prompt guardrails, context segmentation, RBAC enforcement) that raises the difficulty of satisfying that node rather than simply “removing” it from the tree. This explicit placement of defenses, together with the precondition–execution–impact decomposition, is what allows us to reason systematically about where to invest in controls and how multi-step attack paths are reshaped across the LLM-powered healthcare system.

\begin{figure}
    \centering
    \includegraphics[width=1\linewidth]{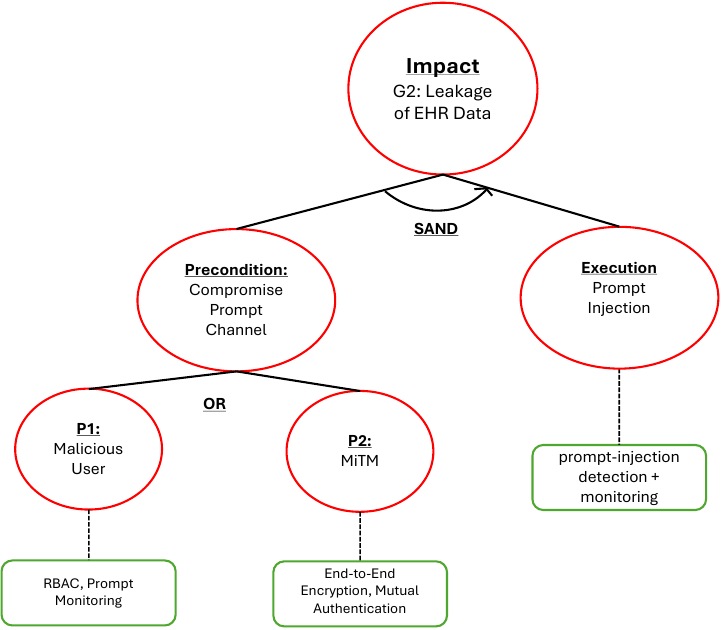}
    \caption{Toy ADT illustrating precondition–execution–impact decomposition and defense placement for G2 (EHR leakage)}
    \label{fig:toy-adt}
\end{figure}

Figure~\ref{fig:toy-adt} shows a simplified ADT for \textit{G2: Leakage of EHR Data} that illustrates how we separate preconditions from execution and how impact is handled. In this toy tree, the attacker must first satisfy a precondition $\mathit{P}$ (compromise the prompt channel), which can occur either via a malicious user or a MitM attack (\texttt{OR}), and then perform the execution step $\mathit{V}$ (prompt injection), modeled as $\mathit{P}\ \texttt{SAND}\ \mathit{V}$. Impact is associated with the goal node (\textit{G2}) and applied when the root is achieved; it is not modeled as an additional sequence of attack steps. Defense nodes attach to specific $\mathit{P}$ or $\mathit{V}$ nodes to increase the difficulty of reaching the goal.

\subsection{CVSS-based Quantification of Attack–Defense Trees}
\label{cvss-method}

We quantify our Attack–Defense Trees using the Common Vulnerability Scoring System (CVSS) ~\cite{FIRST_CVSSv3.1} because it is (i) widely adopted in security operations, (ii) directly tied to real-world vulnerability data, and (iii) decomposes risk into factors that fit naturally onto ADT nodes (attack surface, complexity, required privileges, and user interaction). Alternatives such as DREAD ~\cite{DREAD}, purely qualitative ratings, or ad-hoc likelihood scales are easier to apply but lack standardized scoring and are rarely tied to concrete CVEs, making it difficult to reuse existing vulnerability data or to compare our results with how vulnerabilities are scored and prioritized in practice.

We adopt CVSS v3.1 ~\cite{cvss} as the backbone of our exploitability analysis. 
Because early-stage system analysis often lacks deployment-specific vulnerability evidence, we estimate the feasibility of representative attack steps by mapping them to comparable, publicly scored CVEs and reusing their CVSS v3.1 exploitability vectors; when proof-of-concept exploits or widely reported exploitation patterns exist, they provide additional support for these baseline estimates.
First, the vast majority of vulnerabilities we use as \textit{representative CVEs} for attack steps are still scored in v3.1; using v3.1 lets us reuse their official exploitability vectors without re-estimating metrics. Second, v3.1’s exploitability formulation is stable and well-understood within the security community, which enhances interpretability and reproducibility. Our aggregation method relies solely on the structure of the exploitability sub-score (a multiplicative combination of AV, AC, PR, and UI), so it is straightforward to port to CVSS v4.0 ~\cite{FIRST_CVSSv4.0} in future work by replacing the per-leaf metric weights while retaining the same ADT aggregation logic. We discuss this migration path in the Discussion.

\textbf{Exploitability inside the tree; Impact at the goal.} Inside the ADT, we focus on how easily the attacker can progress, not on the severity of the outcome. Accordingly, all precondition and execution nodes are annotated only with the CVSS exploitability sub-score. At the same time, impact (C, I, A) and scope are reserved for the goal node reached by a completed path. This separation is deliberate: intermediate steps (e.g., session compromise, prompt-channel compromise, or tool misuse) recur across multiple goals, and their feasibility is primarily determined by access, complexity, required privileges, and user interaction. In contrast, severity is goal-specific. Assigning impact to intermediate nodes would double-count consequences along a path and would prevent shared subtrees from being reused consistently across goals. Therefore, we apply impact only once at the goal node reached by a completed path after aggregating exploitability across the path to obtain $E_{\text{path}}$. For example, a shared precondition such as user session hijacking can appear in paths leading to both \textit{G1} and \textit{G2}; its exploitability remains the same, while the impact applied at the goal differs. We use the standard CVSS v3.1 exploitability formula~\cite{cvss}. 

\begin{equation}
E = 8.22 \times AV \times AC \times PR \times UI 
\label{eq:exploitability}
\end{equation}
where each factor is taken from the standard CVSS metric scales: $AV \in \{N,A,L,P\}$, $AC \in \{L,H\}$, $PR \in \{N,L,H\}$ (evaluated under Scope: Unchanged), and $UI \in \{N,R\}$.

\noindent{Metric interpretation:} 
\textit{Attack Vector (AV)} captures the attacker’s required proximity: 
N (Network – remotely exploitable), 
A (Adjacent – same broadcast/domain), 
L (Local – local system access needed), 
P (Physical – physical interaction needed). 
\textit{Attack Complexity (AC)} reflects external conditions needed for exploitation: 
L (Low – straightforward, no special requirements), 
H (High – requires specific timing or configuration). 
\textit{Privileges Required (PR)} indicates initial access level: 
N (None), 
L (Low – regular user), 
H (High – administrative). 
\textit{User Interaction (UI)} depicts dependency on legitimate user actions: 
N (None), 
R (Required – user must click, open, or interact). This design cleanly separates path difficulty (how hard it is to carry out the attack) from goal severity (how damaging it is if the goal is reached). 


We keep CVSS Scope fixed to Unchanged (S:U) in all exploitability calculations. Scope indicates whether a vulnerability in one security authority can affect another; here, all three goals (intervening in medical procedures, EHR leakage, and availability disruption) are defined within the healthcare assistant’s own security domain. Cross-scope effects (e.g., lateral movement into external systems) are out of scope for this study. Fixing S:U keeps the analysis transparent and comparable across goals.

At each goal node, we reintroduce Impact and compute a CVSS-style base score from the path-level exploitability $E_{\text{path}}$ and the chosen impact triple $(C,I,A)$:
\begin{equation}
ISC_{\text{Base}} = 1 - (1-C)(1-I)(1-A)
\label{eq:isc}
\end{equation}
\begin{equation}
\text{Impact} = 6.42 \times ISC_{\text{Base}}
\label{eq:impact}
\end{equation}
\begin{equation}
\text{Base} = \text{round\_up}\left(\min(\text{Impact} + E_{\text{path}}, 10)\right)
\label{eq:base}
\end{equation}
following the standard CVSS v3.1 equations for S:U.

\textbf{How ADT nodes map to CVEs.} Each leaf node in the ADT corresponds to a concrete attack step (e.g., “user hijacking,” “unvalidated shared context,” “task injection”). Following our CVSS grounding approach above, for each leaf ${\ell}$ we select one or more representative CVEs that satisfy two conditions: (i) the attack technique matches the modeled step, and (ii) a CVSS v3.1 scoring vector is publicly documented for that vulnerability. We also ensure that the representative CVE reflects a comparable attacker posture (network, adjacent, local), privilege level (none, low, high), and user-interaction requirement.

From each representative CVE, we extract its CVSS v3.1 exploitability tuple ($(AV_{\ell}, AC_{\ell}, PR_{\ell}, UI_{\ell})$) and compute $E(\ell)$ using (\ref{eq:exploitability}). If multiple CVEs plausibly represent the same attack technique (e.g., several variants of session hijacking or cache-leak exploits), we document the candidate set 
$\mathcal{C}_{\ell}$ and select the worst-case exploitability. When no directly matching CVE exists for an emerging LLM-specific threat, we approximate the metrics by analogical mapping: we identify CVEs with similar attack surface and preconditions (e.g., cache isolation failures, insecure API integrations) and adapt their metric tuples to the LLM scenario, documenting the rationale for transparency.

\textbf{Aggregating exploitability across ADT structure:} $\mathcal{P}$ denotes the set of alternative preconditions for an attack step, and $V$ is the execution step that becomes possible once any precondition in $\mathcal{P}$ is satisfied. Internal ADT nodes use standard OR, AND, and SAND connectors, which we aggregate as follows. \texttt{OR node:}
$E(\mathrm{OR}) = \max_i E(\text{child}_i), $since the attacker chooses the easiest available option. \texttt{AND node:} $E(\mathrm{AND}) = \min_i E(\text{child}_i)$, since the overall difficulty is determined by the hardest required precondition. \texttt{SAND node:} Connecting the precondition family $\mathcal{P}$ to the execution step $V$,
$
E_{\text{path}} = E(\mathrm{SAND}(\mathcal{P} \to V))
               = \min\big(E(\mathcal{P}),\, E(V^\star)\big),
~\label{eq:sand}
$
where $E(\mathcal{P})$ is the exploitability of satisfying any precondition in the family,
and $E(V^\star)$ is the exploitability of executing $V$ after those preconditions are met (defined next).

Within a precondition family 
$\mathcal{P} = \{p_1,\ldots,p_k\}$, we first compute its exploitability as an OR-node: $E(\mathcal{P}) = \max_i E(p_i)$, because an attacker only needs to succeed with the easiest available precondition to proceed. However, we also need to capture how the overall conditions established by these preconditions change the difficulty of the subsequent execution step $V$. CVSS isolates this notion in the Attack Complexity (AC) metric. Intuitively, different preconditions can create different ``operating environments'' for the attacker: some yield a stable, repeatable foothold (making the next step straightforward), while others require fragile circumstances (e.g., rare timing or system state), making the next step harder even if the attacker targets the same interface. We model this dependency by conditioning only the attack complexity ($AC$) of $\mathit{V}$ on the precondition family. 
For each precondition $p_i$, we record its complexity label $AC(p_i) \in \{L, H\}$ from the mapped CVE. We then compute a majority Attack 
Complexity:
\begin{equation}
AC_{\text{maj}} =
\begin{cases}
L, & \text{if more preconditions in } P \text{ have } AC = L, \\
H, & \text{otherwise (ties default to } H \text{, conservative).}
\end{cases}
\label{eq:majority_ac}
\end{equation}

Intuitively, if most viable ways to enable the execution step are low-complexity attacks (e.g., commodity phishing or straightforward misconfiguration abuse), then operating within this prepared environment also tends to be low complexity. Conversely, if high-complexity preconditions dominate (e.g., precise cache manipulation or rare timing conditions), we treat the subsequent step as high complexity. We deliberately propagate only AC, not the full exploitability score, to avoid double-counting other factors such as privileges or user interaction, which belong to the specific execution step. More precisely, $AV$, $PR$, and $UI$ are properties of the execution interface and requirements encoded by the representative CVE for $\mathit{V}$; propagating them from $\mathit{P}$ into $\mathit{V}$ would count the same access, privilege, and user-dependence twice---once in reaching the precondition family and again in executing $\mathit{V}$. Conditioning only $AC$ captures the intended dependency (``how constrained the attacker is once the environment is prepared'') without duplicating step-specific requirements.

We then construct the execution node vector $V^\star$ by overriding only the AC component of $V$’s CVSS tuple and keeping its $AV$, $PR$, and $UI$ values from the representative CVE: $E(V^\star) = 8.22 \times AV_V \times AC_{\text{maj}} \times PR_V \times UI_V.
\label{eq:vstar} $

This yields a clean separation of responsibilities:
(i) $E(P)$ captures how easy it is for the attacker to establish the preconditions under which $V$ becomes feasible;
(ii) $AC_{\text{maj}}$ captures how the environment created by those preconditions alters the difficulty of executing $V$; and
(iii) the SAND operator combines these into a path-level exploitability: we combine them as defined in Eq (~\ref{eq:sand}).  which is then propagated upward through the remainder of the ADT using the OR/AND aggregation rules.

\textbf{Goal-Level Impact and Final CVSS Score.}
Once $E_{\text{path}}$ has been propagated from the ADT using the OR/AND/SAND rules and Majority-AC conditioning, we introduce impact only at the goal node. For each goal $G_j \in \{G_1, G_2, G_3\}$, we specify a triple $(C_j, I_j, A_j)$ that reflects the severity of achieving that goal in the healthcare setting. For example, for Prompt Injection under $G_1$, we set $C = 0.00$, $I = 0.56$, $A = 0.00$, representing a primarily integrity-driven hazard. Using these values, we compute the impact sub-score exactly as in CVSS v3.1 using Eq (~\ref{eq:isc})(~\ref{eq:impact}). We then combine this with the path-level exploitability to obtain the CVSS base score using Eq (~\ref{eq:base}). This convention ensures two key properties:(i) Different goals that share ADT subtrees (e.g., user session hijacking under both $G_1$ and $G_2$) reuse the same exploitability structure.
(ii)Differences in risk severity across goals are cleanly expressed through the assigned $(C, I, A)$ impact values, without entangling severity into the internal nodes of the tree.

\begin{figure}
    \centering
    \includegraphics[width=1\linewidth]{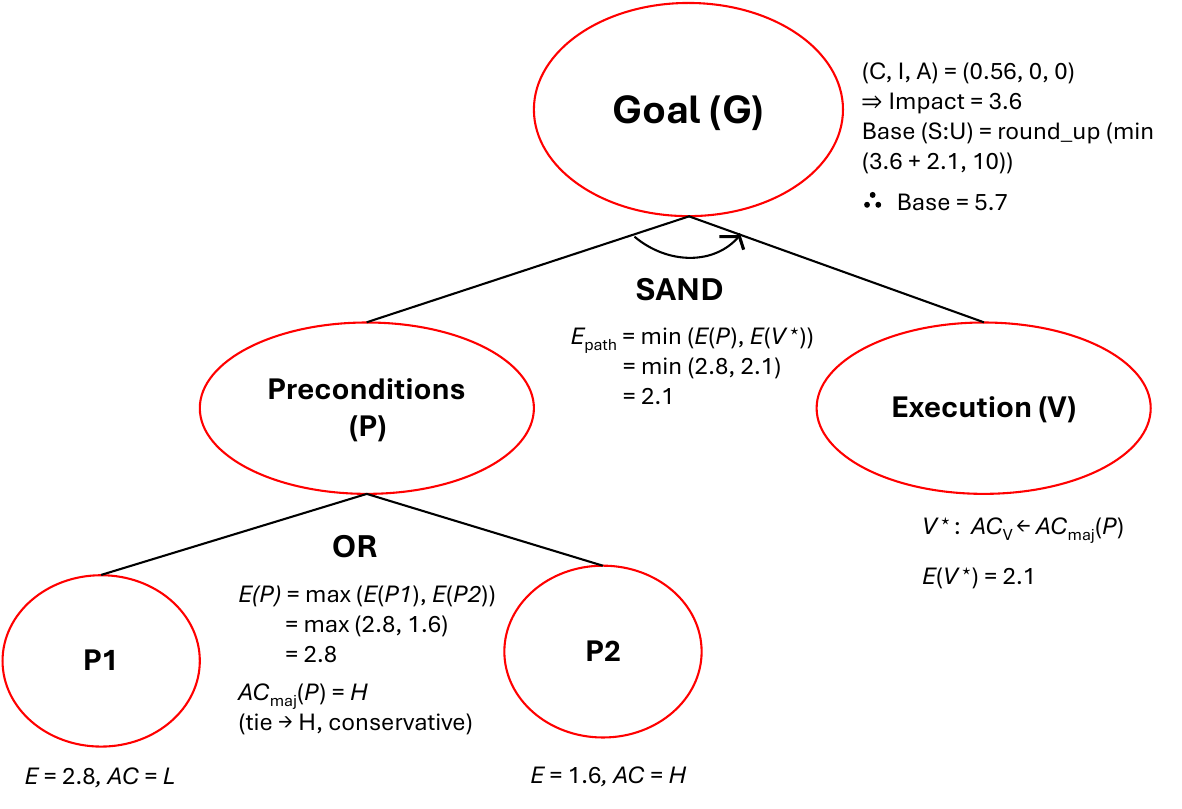}
    \caption{Toy example showing how we compute a path-level CVSS-style score}
    \label{fig:exploitability_example}
\end{figure}

Consider a simplified path (Refer Figure ~\ref{fig:exploitability_example}) $\mathit{P}\ \texttt{SAND}\ \mathit{V}$ with two alternative preconditions $\mathit{P}=\{p_1,p_2\}$, where the mapped CVEs yield $E(p_1)=2.8$ with $AC(p_1)=L$ and $E(p_2)=1.6$ with $AC(p_2)=H$, giving $E(\mathit{P})=\max\{2.8,1.6\}=2.8$; with a tie-free majority $AC_{\mathrm{maj}}=H$, we compute $E(\mathit{V}^{\star})$ by replacing only $AC_{\mathit{V}}$ with $AC_{\mathrm{maj}}$ in $\mathit{V}$'s CVSS tuple (Eq.~\ref{eq:vstar}), assume $E(\mathit{V}^{\star})=2.1$, and obtain the path exploitability $E_{\text{path}}=\min\{2.8,2.1\}=2.1$ (Eq.~\ref{eq:sand}); finally, applying goal-level impact once at the root for $(C,I,A)=(0.56,0,0)$ yields $ISC_{\text{Base}}=0.56$ (Eq.~\ref{eq:isc}), $\mathrm{Impact}=6.42\times0.56=3.60$ (Eq.~\ref{eq:impact}), and $\mathrm{Base}  =\mathrm{round\_up}(\min(3.60+2.10,10))=\mathrm{round\_up}(5.70)$ (Eq.~\ref{eq:base}), illustrating that exploitability accumulates along the attack sequence while impact is introduced only once at the goal node.

\subsection{Risk Treatment}
\label{treat-method}

\textbf{Risk treatment via exploitability reduction.}
Our risk treatment focuses on reducing path-level exploitability while keeping the goal-level impact parameters $(C, I, A)$ fixed, as defined in the CVSS section. For any attack path $X$ that we model as a SAND composition of a precondition family $P$ and an execution step $V$ (i.e., $X = \mathrm{SAND}(P \to V)$), we propagate exploitability using Eq.~(\ref{eq:sand})
where $E(P)$ is the OR-aggregated exploitability of the precondition family and $E(V^\star)$ is the execution-step exploitability after conditioning on the (possibly hardened) preconditions (Eq.~(\ref{eq:vstar})).

\textbf{How defenses change $E$.} \textit{Leaf transforms:} Each defensive control is modeled as a transformation on one or more CVSS base metrics at specific leaves of the ADT. For a leaf $\ell$ with baseline vector $(AV_\ell, AC_\ell, PR_\ell, UI_\ell)$, a control may, for example:
(i) raise the privileges required (e.g., MFA: $PR{:}\ \text{L} \rightarrow \text{H}$);
(ii) increase attack complexity (e.g., strong parsing or isolation: $AC{:}\ \text{L} \rightarrow \text{H}$); or
(iii) introduce a human gate (e.g., HITL review: $UI{:}\ \text{N} \rightarrow \text{R}$). After applying all orthogonal transforms that protect a leaf $\ell$, we recompute its exploitability $E(\ell)$ using Eq.~(\ref{eq:exploitability}), with the same CVSS v3.1 weights as in our scoring section.

\textit{Exporting hardened preconditions to the execution step:} Conditioning the system on a hardened set of preconditions $P$ also changes the practical difficulty of the execution step $V$. After applying defenses on $P$, we recompute both the aggregate exploitability $E(P)$ (Eq.~(\ref{eq:exploitability})) and the majority Attack Complexity label $AC_{\text{maj}}(P)$ derived from the precondition CVEs (Eq.~(\ref{eq:majority_ac})). If no explicit execution-level controls are deployed, we instantiate the conditioned execution vector $V^\star$ by replacing the original $AC_V$ with $AC_{\text{maj}}(P)$, while keeping $AV_V$, $PR_V$, and $UI_V$ from $V$’s representative CVE, and compute $E(V^\star)$ via Eq.~(\ref{eq:vstar}).

\begin{figure}
    \centering
    \includegraphics[width=1\linewidth]{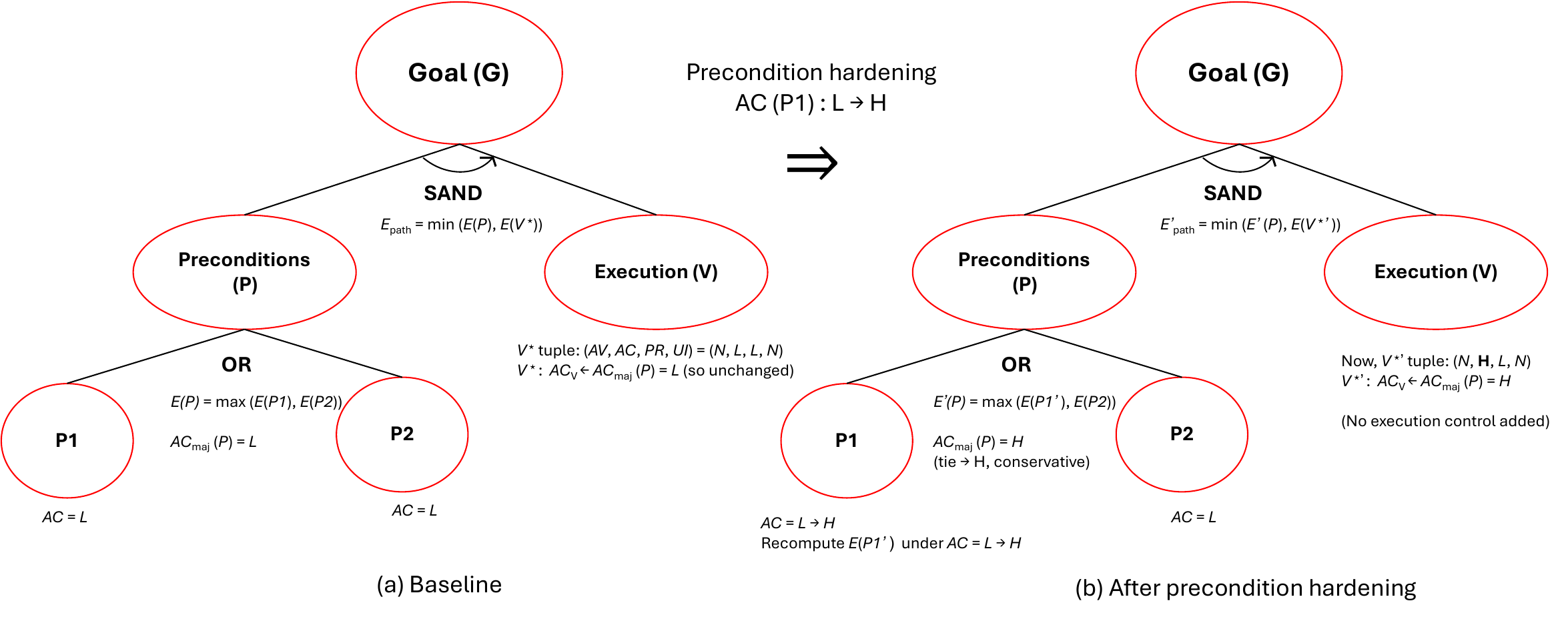}
    \caption{Toy example for risk treatment}
    \label{fig:risk_treatmeant_example}
\end{figure}

Consider a path (Refer Figure ~\ref{fig:risk_treatmeant_example}) $X=\texttt{SAND}(\mathit{P}\rightarrow \mathit{V})$ where $\mathit{P}=\{p_1,p_2\}$ are alternative preconditions and $\mathit{V}$ is an injection-style execution step. Suppose initially $AC(p_1)=L$ and $AC(p_2)=L$, so $AC_{\mathrm{maj}}(\mathit{P})=L$. If $\mathit{V}$'s representative CVE has $(AV_{\mathit{V}},AC_{\mathit{V}},PR_{\mathit{V}},UI_{\mathit{V}})=(N,L,L,N)$, then $\mathit{V}^{\star}$ remains $(N,L,L,N)$ and $E(\mathit{V}^{\star})=E(\mathit{V})$. Now assume we deploy a precondition-hardening control on $p_1$ that raises its complexity $AC(p_1):L\rightarrow H$ (e.g., stronger isolation or stricter policy checks), while $p_2$ remains $L$. Then the majority complexity becomes $AC_{\mathrm{maj}}(\mathit{P})=H$ (ties default to $H$, conservative), and we export this to the execution step by setting $\mathit{V}^{\star}=(AV_{\mathit{V}},AC_{\mathrm{maj}}(\mathit{P}),PR_{\mathit{V}},UI_{\mathit{V}})=(N,H,L,N)$. This reduces $E(\mathit{V}^{\star})$ relative to $E(\mathit{V})$ even though no execution-level control was added, capturing that a harder-to-achieve foothold typically yields a less permissive environment for exploitation. The updated path exploitability is then $E_{\text{path}}=\min(E(\mathit{P}),E(\mathit{V}^{\star}))$ (Eq.~\ref{eq:sand}).

\textbf{What defenses do not change  E.} We explicitly distinguish between controls that change the CVSS base vector and controls that primarily affect \textit{post-compromise outcomes}. Detective and operational mechanisms such as anomaly alerts, key/certificate rotation frequency, logging, and post-hoc audits do reduce overall risk by improving detection and response (e.g., shortening time-to-detect and limiting attacker dwell time, blast radius, and realized loss). However, these mechanisms typically do not increase the attacker’s effort to achieve a single successful exploit and therefore do not directly alter the CVSS base metrics ($AV$, $AC$, $PR$, $UI$) used to compute exploitability. Accordingly, we model such controls outside the exploitability score $E$: they are discussed as complementary risk-reduction measures, but they leave $E$ unchanged in our CVSS-based treatment. This separation keeps the exploitability computations interpretable and avoids mixing attacker effort with downstream response-driven loss reduction.

\textbf{Canonical Treatment Scenarios.} For each attack path, we do not try to optimize over an exponential defense space. Instead, we analyze four canonical “design points” that map directly to realistic engineering choices: 1. Harden major preconditions only. Identify the dominant preconditions in (P) (those with the highest (E) and/or lowest AC) and apply a small number of controls that substantially increase their difficulty. 2. Harden all preconditions. Apply selected controls to every leaf in (P), raising the overall bar to reach the execution step. This represents an “infrastructure-first” strategy. 3. Harden execution only. Leave (P) unchanged, but add guardrails on (V) (e.g., canonicalization, task allowlists, HITL gates) that increase (AC) and/or (UI) for the execution node. This corresponds to a “guardrails-first” approach. 4. Harden both sides. Combine strategies (2) and (3), simultaneously raising the cost of satisfying preconditions and of turning those footholds into a successful exploit. These scenarios span the main trade-off surface without requiring exhaustive search over all control combinations.

\textbf{Computation Workflow.} For each path $X$: 1. Select the scenario and corresponding defense set. 2. Apply transforms to affected CVSS vectors on $P$ and/or $V$. 3. Recompute $E(P)$ using OR aggregation across preconditions. 4. Determine $AC_{\text{maj}}(P)$ and export it to the execution step. 5. Compute $E(V^\star)$ using the exported or transformed vectors. 6. Propagate via SAND. 7. Report updated $E_{\text{path}}$, $\Delta E$, and total cost level.

\textbf{Cost Model.} To compare defenses across goals and scenarios, we assign each control an ordinal cost level from 1 to 4 that captures four dimensions: Engineering effort, Platform/infrastructure footprint, Operations load, and UX/process impact. We do not attempt to convert these into monetary values; instead, they function as a ranking scale that can be mapped to local budgets or staffing estimates. The levels refer to Table~\ref {tab:cost_levels} For example, adding an application-level allowlist for tool calls or tightening RBAC policies is typically Level~1-2 (localized configuration and limited code changes), whereas deploying mutual TLS end-to-end with certificate lifecycle management and continuous monitoring is typically Level~3 (cross-service integration, PKI and secrets management, and sustained operational load).

\begin{table*}[!t]
\centering
\renewcommand{\arraystretch}{1.2}
\setlength{\tabcolsep}{3pt} 

\begin{tabularx}{\textwidth}{|c|l|X|X|X|X|}
\hline
\textbf{Level} & \textbf{Label} & \textbf{Engineering effort} &
\textbf{Infra / tooling footprint} & \textbf{Operations load} &
\textbf{UX / process impact} \\
\hline
1 & Very Low &
Minor policy/config; single engineer; no core-path code changes &
Reuse existing services; no new components &
“Set-and-forget” checks; negligible monitoring &
Invisible to users \\
\hline
2 & Low &
Small project; limited code changes; one team &
Add a light service/lib; standard CI/CD integration &
Periodic checks/rotations; light on-call playbooks &
Minor extra step for some users \\
\hline
3 & Medium &
Cross-team work; non-trivial refactors; integration/testing &
Multiple services or control plane; new secrets/PKI &
Continuous upkeep (dashboards, tuning); measurable SOC workload &
Noticeable friction/training; workflow changes \\
\hline
4 & High &
Org-wide program; architectural changes; specialized skills &
New platform class or hardware; PKI re-plumb; broad dependency changes &
Dedicated runbooks/audits; sustained operators; compliance reporting &
Significant user/process change; likely perf/latency trade-offs \\
\hline
\end{tabularx}
\caption{Ordinal Cost Levels for Defense Measures}
\label{tab:cost_levels}
\end{table*}

%% file: sections/results.tex
\section{Results}

\subsection{RQ1 – Goal-Driven Attack–Defense Trees}

We realize our methodology by building three goal-oriented ADTs G1 (intervening in medical procedures), G2 (EHR data leakage), and G3 (disruption of access or availability) ~\cite{healthsec2026}. Each tree factors the goal into ~\cite{healthsec2026}: (i) preconditions, which capture the adversary footholds tied to specific components and trust boundaries; (ii) execution steps, which encode empirically observable attacker behaviors; and (iii) a final-impact node representing the security goal. Defense nodes attach to preconditions and execution steps at feasible control points, but we defer a detailed discussion of these defenses to our risk treatment section. In this part of the paper, we focus on constructing the attack paths themselves, turning the STRIDE-derived threat taxonomy from our prior work into explicit, multi-step attack scenarios

\subsubsection{G1: Intervening in medical procedures}

This goal reflects adversaries manipulating LLMs in healthcare systems to disrupt medical decision-making and cause harmful outcomes. Figure~\ref{adt_g1} presents the corresponding Attack Tree. The root node, "Intervene in Medical Procedures," branches into five primary attack paths, each with its preconditions and actions for exploitation: (1) Prompt Injection, (2) LLM Session Management, (3) Orchestration Errors, (4) Model Tampering, (6) MiTM Web Session. We elaborate on each path as follows ~\cite{healthsec2026}: 

\textbf{Prompt injection}. Successful execution typically involves two distinct stages: first, establishing unauthorized control over prompt delivery channels (precondition), followed by injecting crafted inputs that exploit LLM behavior (attack execution).

\textit{Preconditions}: Compromise of Prompt Channel - Prompt injection depends on gaining control over the system’s input pathway. Any one of the following (modeled as OR conditions) provides sufficient access: 1. User Hijacking: Acquiring legitimate user credentials or session tokens through phishing or credential theft to manipulate inputs directly. 2. User Machine Hijacking: Compromising end-user devices via malware to intercept or alter prompts before submission. 3. Malicious User: Authorized users deliberately inputting adversarial content. 4. Stolen Session API Keys: Exploiting exposed or misconfigured API keys to inject prompts directly at the backend. 5. Man-in-the-Middle (MitM): Intercepting and modifying prompts in transit through compromised or unsecured network channels. These preconditions establish the necessary presence or influence within the prompt-delivery chain.

\textit{Attack Execution}: Crafting Suitable Prompts - Once access is gained, adversaries deliver harmful instructions through either direct or indirect injection: 1. Direct Prompt Injection ~\cite{Rossi2024}: Injecting adversarial content into the LLM input stream through: a. Human-Based Injection: Explicitly written instructions. b. Parameter-Based Injection: Malicious commands embedded in API parameters or HTTP requests. c. Obfuscation-Based Injection: Concealed payloads using encoding, character substitutions, invisible tokens, or visually embedded instructions in images ~\cite{fllm_2025} or documents. d. Optimization-Based Injection: Leveraging the LLM’s predictive tendencies to guide it toward harmful completions. 2. Indirect Prompt Injection ~\cite{Greshake}: Manipulating trusted intermediate systems (e.g., translators, knowledge bases, agent tools) to relay hidden instructions. This requires the AND precondition: a. Hidden Instruction Integration - Malicious instructions must be covertly embedded within benign-looking textual or visual ~\cite{fllm_2025} external content, ensuring they remain unnoticed yet executable by the LLM. Once met, indirect injection proceeds via: Translator Tampering: Altering translation outputs. Knowledge Base (KB) Poisoning ~\cite{chen2024agentpoison}: Inserting adversarial entries into external databases. Social Engineering: Convincing authorized users or tools to unknowingly relay compromised content OR b. Agent Tool Injection: Introducing harmful instructions through compromised plugins or integrated tools.

\begin{figure}
    \centering
    \includegraphics[width=1\linewidth]{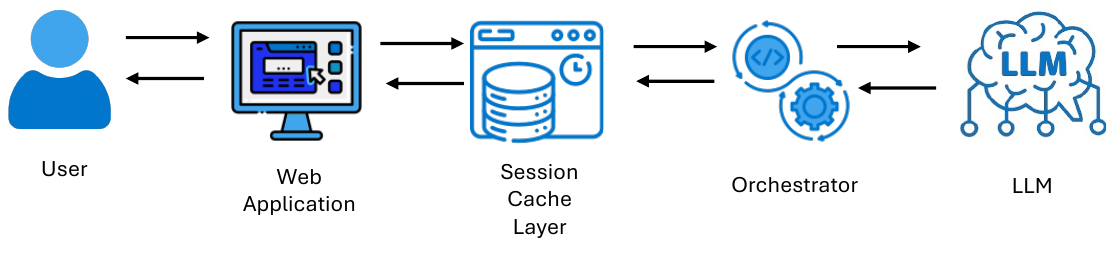}
    \caption{Workflow of the LLM-based Healthcare System.}
    \label{workflow}
\end{figure}

\textbf{LLM Session Mismanagement}. An LLM session in our architecture is a stateful interaction between the user-facing web application and the LLM, coordinated by the orchestrator agent. It encompasses prompt–response exchanges, retained conversational context, and per-interaction metadata used to track task progress. As shown in Figure~\ref{workflow} ~\cite{healthsec2026}, the orchestrator preserves continuity by relaying user inputs to the LLM, managing intermediate results, and maintaining short-term memory across steps to support multi-turn reasoning and consistent clinical outputs. Session mismanagement occurs when controls such as strong authentication, user/session isolation, and context clearing are insufficient, enabling an attacker to interfere with or carry state across sessions.

\textit{Preconditions} – Weak Session Handling To exploit the LLM session layer, the attacker must first take advantage of systemic vulnerabilities in how sessions are managed. In our tree, this is modeled as an OR condition- 1. Weak Authentication: Session tokens or credentials are accepted without rigorous verification, allowing spoofing or the use of expired sessions. 2. Lack of Isolation: User contexts are not strictly separated, enabling cross-session memory contamination or influence over shared prompt history. 4. No Context Purging: Dialogue turns or task metadata are not cleared at session end, letting an adversary inherit or reuse contextual content later. These conditions allow an adversary to shape LLM behaviour across requests and to jeopardize output integrity.

\textit{Attack Execution:} Session Takeover; 1. Session Hijacking: Taking control of an active session by intercepting tokens, exploiting insecure storage, or reusing credentials, then injecting prompts, altering reasoning, or corrupting retained memory. 2. Session Fixation: Planting a pre-chosen session identifier and tricking the orchestrator into accepting it, granting persistent access as the legitimate user interacts, and enabling subtle prompt manipulation or context poisoning. Both paths undermine temporal boundaries and logical consistency in LLM exchanges; compromised memory can quietly redirect clinical instructions, overwrite diagnostic logic, or disable safety guardrails.

\textbf{Orchestration Errors}. Orchestration Errors arise when attackers manipulate the control logic coordinating task execution, data routing, or interactions with the LLM. Such manipulation can redirect, alter, or inject operations, allowing adversaries to influence how the LLM interprets inputs and generates responses, even without direct access to the model itself. These vulnerabilities pose serious risks to the integrity of downstream clinical decisions.

\textit{Preconditions}: Compromising the Orchestration Workflow ; To exploit orchestration logic, attackers must first compromise the workflow's integrity. This attack precondition is satisfied through any of the following weaknesses (modeled as an OR condition): 1. Lack of Integrity Checks: Absence of verification mechanisms for control flows and intermediate task results allows forged or tampered instructions to be accepted by the orchestrator unchecked. 2. Weak Authentication: Orchestration APIs or internal task management components may be exposed without adequate credential enforcement, permitting unauthorized access or manipulation. 3. Unvalidated Shared Context Memory: Intermediate memory (e.g., conversation state, task metadata) is not adequately validated before reuse across multiple stages. This permits adversaries to poison or overwrite context to influence task planning or execution logic. These weaknesses enable attackers to inject malicious control signals, alter the sequence of task execution, or redirect workflows.

\textit{Attack Execution}: Orchestration Manipulation; Once the orchestrator is compromised, attackers can manipulate it in several ways, as formalized in our tree: 1. Task Injection: Attackers introduce unauthorized or adversarial tasks into the orchestrator’s workflow. These tasks may request sensitive data, alter LLM output constraints, or force the invocation of unsafe external tools. 2. Task Mismanagement: Bugs, logic flaws, or adversarial manipulation may cause the orchestrator to misroute tasks to the wrong agents or process them in an incorrect order. This leads to context mismatch, contradictory outputs, or inappropriate medical recommendations. 3. Memory Poisoning ~\cite{chen2024agentpoison}: Malicious inputs are injected into intermediate orchestrator memory (e.g., shared variables, pipeline buffers). If reused by downstream components without sanitization, they can silently corrupt the decision logic or mislead the LLM. 4. Agent-in-the-Middle Attacks ~\cite{he2025}: In orchestrator frameworks that use external tool agents or plugins (e.g., medical database lookups or translation modules), attackers may replace or spoof these agents, injecting falsified data or responses into the pipeline. These attacks exploit the orchestrator’s centrality and authority over the system’s task flow, often resulting in silent and persistent degradation of clinical accuracy.

\textbf{Model tampering}. involves the unauthorized manipulation, replacement, or extraction of a deployed model to alter its behavior or leak knowledge. In a healthcare context, this can corrupt diagnosis logic, introduce bias, and degrade decision reliability. Our tree models three primary methods under this vector: Poisoning, which compromises the model’s training or fine-tuning data. Replacement: Substituting the legitimate model with a malicious one. Extraction: Exploiting model access to steal proprietary parameters or training data.

\textit{Preconditions}: Model Tampering Access;
Before adversaries can perform model tampering, certain conditions must be met, grouped under the Preconditions-MT subtree. These include: 1. Gaining Access to Model Artifacts: Attackers must acquire access to the model's binary weights, architecture, or intermediate outputs, often through misconfigured storage, overly permissive roles, or exposed APIs. 2. Lack of Integrity Verification: In the absence of digital signatures, attestation mechanisms, or checksum-based validation, model files can be silently replaced or modified during deployment or runtime.
Together, these vulnerabilities enable an adversary to replace or manipulate a model undetected, laying the groundwork for more targeted tampering strategies.

\textit{Attack Execution}; Once the above preconditions are met, the attacker may proceed with one of the following three classes of attacks, as detailed in our tree. 1. Model Poisoning
Model poisoning ~\cite{OWASP2025} refers to the adversarial manipulation of the model’s parameters or decision boundaries, typically by injecting malicious data or perturbations into the training pipeline. This attack path is further conditioned on the Precondition-MP branch: Access to the Training Pipeline or Data, Model Retraining or Fine-tuning is Scheduled, or Lack of Data Sanitization or Filtering. When these conditions are satisfied, attackers may introduce: a. Data Poisoning: Corrupt training data injected into the fine-tuning stage alters model behavior, leading to incorrect diagnoses or manipulated output for specific inputs. b. Targeted Poisoning: The attacker implants backdoors into the model that activate only when specific trigger phrases or patient attributes are encountered, making the tampering stealthy and targeted. 2. Model Replacement refers to the complete substitution of the deployed LLM with a compromised or malicious variant. In this case, the attacker may deploy a model that mimics the original’s behavior on most inputs but behaves maliciously under specific clinical contexts. Without strong version attestation and deployment control, such substitution can remain undetected, especially in containerized or multi-agent environments. 3. Model Extraction: Model extraction attacks aim to steal proprietary model parameters or replicate model behavior through repeated querying ~\cite{carlini2024steal}.

\textbf{MiTM Web Session} attack occurs when an adversary intercepts and potentially modifies communication between the user and the web application interface. In the context of LLM-powered healthcare systems, this enables the attacker to inject, suppress, or alter user queries before they reach the orchestrator or LLM, thereby silently manipulating the clinical response workflow. As this conventional attack is widely known, we do not elaborate on this attack path in this study.

\subsubsection{G2: Leakage of Electronic Health Record (EHR) Data}

This goal captures threats leading to unauthorized access, inference, or exfiltration of sensitive Electronic Health Record (EHR) data. Such leakage compromises patient privacy, erodes trust, and violates regulatory frameworks like HIPAA. In LLM-based systems, features such as session memory, external tool integration, and cached inference increase the risk of both accidental and adversarial disclosure.

Figure~\ref{adt_g2} presents the Attack Tree constructed for G2, detailing the major threat paths and the preconditions that enable attackers to compromise data confidentiality.

\textbf{User Session Compromise} leads to EHR leakage via two main vectors. First, man-in-the-middle (MitM) attacks allow interception of prompts and LLM responses containing sensitive data. Second, with session access, adversaries can use prompt injection to extract prior context or inferred patient information. While structurally modeled in G1, the emphasis here is on confidentiality loss rather than care disruption.

\textbf{LLM Session Compromise} occurs when unauthorized access to memory or cache states within multi-turn LLM sessions leads to leakage of sensitive EHR data. As modeled in our AT, this includes session hijacking, fixation, and cache compromise.

\textit{Preconditions:} These attacks emerge when session isolation is weak, memory is not purged between sessions, sandboxing is incomplete, or authentication is insufficient.

\textit{Attack Execution:} 1. Session Hijacking: An attacker takes over an active session to retrieve previous EHR-related context. 2. Session Fixation: The attacker tricks the system into using a known session ID, gaining access to the victim’s memory. 3. Cache Compromise: Modern LLM inference often uses key-value (KV) caches to optimize for latency, especially in systems with high query throughput. If cache boundaries are poorly enforced, attackers can extract tokens or responses from past sessions. Two main mechanisms are modeled: a. Cross-User Cache Exposure: If the system uses a shared cache pool (e.g., Redis) ~\cite{OpenAI-ChatGPT-Outage-2024}, and response-to-user validation is missing or faulty, an adversary can receive cached completions associated with another user’s query. This often stems from architectural shortcuts or improper tagging of cache entries. b. KV-Cache Leakage (PromptPeek-style) ~\cite{DBLP:conf/ndss/WuZZWNWZ25}: In transformer-based architectures, model-level KV caches ~\cite{DBLP:conf/ndss/WuZZWNWZ25} may retain query embeddings across users in multi-tenant pipelines. If no per-session isolation is implemented, and the model pipeline lacks query validation or filtering, an attacker can extract token completions using probing inputs. Recent work such as PromptPeek ~\cite{DBLP:conf/ndss/WuZZWNWZ25} demonstrates this leakage is practical under these conditions.

\textbf{Training Data Leakage} ~\cite{274574} refers to adversarial techniques that extract sensitive EHR content memorized by the LLM during training or fine-tuning. These attacks do not rely on session compromise but instead exploit the model’s internal representations to reconstruct or infer protected patient data. This becomes especially dangerous when the model is trained on real or poorly anonymized EHRs, as even partial leakage can result in serious privacy violations under healthcare regulations like HIPAA.

\textit{Preconditions:} This threat arises when:1. Memorization: the model retains sensitive EHR data due to overfitting or inadequate anonymization; 2. Accessible Query Interface: attackers can query the model directly; 3. Interface Misconfigurations: lack of rate limiting or output filtering enables large-scale probing and exposure of raw completions.

\textit{Attack Execution:} Training Data Extraction involves crafting targeted prompts to reconstruct exact content, such as EHR notes or identifiers, from the training set. Model Inversion infers likely patient attributes, such as conditions or demographics, based on outputs generated from known or partial inputs. While both attacks exploit memorized or internalized data, extraction focuses on recovering exact content, whereas inversion reconstructs generalized patterns.

These attacks are particularly concerning in LLM-based healthcare systems because they target the model’s internal behavior rather than its session or orchestration logic. As a result, they enable passive data harvesting and can bypass traditional access controls, especially in public or weakly monitored deployments.

\textbf{Membership Inference Attack} is modeled as a parallel attack path. Here, the attacker determines whether a specific patient’s data was included in the model’s training set ~\cite{carlini2022membership}. Even without reconstructing content, this alone can breach confidentiality and violate regulations like HIPAA.

\subsubsection{G3: Disruption of Access to EHR Data}

This goal models adversarial threats that prevent legitimate users—patients, clinicians, or support systems—from accessing Electronic Health Record (EHR) data in a timely or reliable manner. While confidentiality and integrity may remain intact, such attacks degrade availability and clinical usability, especially in time sensitive scenarios like emergency care. In LLM-based systems, disruptions can stem from both traditional denial-of-service (DoS) vectors and LLM specific misuse patterns.

We constructed an Attack Tree (AT) for G3 to capture the key preconditions and attack paths that lead to access disruption. Figure~\ref{adt_g3} summarizes the structure, which includes both conventional and LLM specific overload strategies.

\textit{Preconditions:} These attacks require missing or misconfigured controls, including: 1. No Query Rate Limiting: allows adversaries to flood the system with prompts or API calls. 2. Poor Session Management: unused sessions accumulate due to lack of timeout or cleanup. 3. Poor Authentication: unauthenticated users or spoofed agents overwhelm resources. 4. No Input Validation: recursive or malformed prompts pass unchecked. 5. Unthrottled Input APIs – plugin endpoints accept unregulated input streams.

\textit{Attack Execution:} Once preconditions hold, access may be disrupted via: 1. Prompt Flooding ~\cite{ATLAS2025}: Overwhelms the LLM with high-volume or complex prompts. Subtypes include: Excessive Prompt Submission: Standard queries submitted repeatedly to deplete throughput. DoS via Prompt Chaining: Recursive instructions cause nested reasoning loops. These attacks are potent when query costs scale with prompt complexity and no throttling exists. 2. Orchestrator Abuse: Exploits weak task logic. Infinite Agent Loops trigger unresolved cycles between tools and agents. Resource Starvation Attacks tie up compute/memory via sleep commands or long loops. 3. Resource Exhaustion: Directly depletes memory, compute, or task queues. Memory Bombing uses large prompts to trigger token overflow or context spills. Pipeline Queue Flooding submits concurrent tasks to overwhelm orchestration. While both starvation and exhaustion degrade availability, the former deprives specific internal processes, while the latter overloads system-wide resources. 4. External Tool/API Dependency LLM pipelines often invoke third-party services (e.g., translators). If these calls lack protections, Man-in-the-Middle (MitM) attacks can intercept or tamper with responses. Additionally, attackers may launch DoS attacks on those external APIs, preventing LLMs from receiving required outputs and silently stalling pipeline execution. 5. Prompt Injection: Malicious prompts trigger excessive reasoning, recursion, or internal task chains. Prompt chaining differs in that its primary aim is to overwhelm resources through recursive logic, whereas prompt injection hijacks model behavior (e.g., bypassing filters). Both can induce infinite loops and recursive calls, but their intent and impact differ. 6. Model Tampering: Poisoned Fine-Tuning introduces inefficiencies during retraining. Backdoor Positioning causes intentional delays or errors when triggered by specific inputs. 7. Access Control Abuse:
Weak controls may let unauthorized users bypass limits, opening privileged execution paths that exhaust resources.

EHR access disruption poses a serious operational risk in LLM-powered healthcare systems. These attacks are often indirect and subtle, making early detection difficult and impact potentially severe in real-world deployments.

\subsection{Quantitative ADT Analysis using CVSS Exploitability}

To address RQ2, we apply the scoring procedure from Section~3 to representative attack paths. We present the full CVSS computation for the $G_1$ prompt-injection branch (B1) as an end-to-end example, and then summarize the resulting path exploitability scores for the remaining $G_1$ branches as well as the main trees under $G_2$ and $G_3$ using the same workflow.

\textbf{G1  Intervening in Medical Procedures.}
$G_1$ captures how adversaries can manipulate the LLM-based assistant to alter diagnoses, treatment suggestions, or clinical workflows. The corresponding ADT decomposes this goal into five top-level branches: B1 Prompt Injection, B2 LLM Session Mismanagement, B3 Orchestration Errors, B4 Model Tampering, and B5 Web MiTM. We quantify each branch using the CVSS v3.1 exploitability model from Section~3 and attach an integrity-focused impact at the goal.

\textbf{B1: Prompt Injection.}
Prompt injection manipulates LLM inputs to steer clinical recommendations. In our ADT, this branch is modeled as a SAND of preconditions $P$ (compromise of the prompt channel) and execution $V$ (crafting and delivering malicious prompts).

\textit{Preconditions – Compromise of the Prompt Channel.}
An attacker needs one of five preconditions to control what reaches the LLM (OR-node). We selected some representative vulnerabilities from the relevant databases to approximate the risk score of each path. For this purpose, each leaf is mapped to a representative CVE and CVSS v3.1 exploitability vector: \textit{User hijacking} ($\rightarrow$ CVE-2025-22222, AV:N, AC:L, PR:L, UI:N, $E=2.84$), \textit{User machine hijack} ($\rightarrow$ CVE-2025-22457/53770/32433, AV:N, AC:L, PR:N, UI:N, $E=3.89$), \textit{Malicious user} ($\rightarrow$ CVE-2025-42957, AV:N, AC:L, PR:L, UI:N, $E=2.84$), \textit{Stolen API key} ($\rightarrow$ CVE-2024-12012, AV:N, AC:L, PR:L, UI:R, $E=2.07$), and \textit{MiTM on web session} ($\rightarrow$ CVE-2024-50691, AV:N, AC:H, PR:N, UI:N, $E=2.22$).
Taking the OR over these leaves yields $E(P)=\max\{3.89,\,2.84,\,2.84,\,2.07,\,2.22\}=3.89$. The corresponding $AC$ labels are $(L,L,L,L,H)$, so the majority is low-complexity: $AC_{\text{maj}}(P)=L$.

\textit{Execution – direct and indirect injection.}
Conditioned on $P$, the attacker can inject prompts either directly (human-entered, parameter-based, obfuscated, or optimized prompts) or indirectly (via translators, knowledge bases, or agent tools). We represent these as an OR of two children: \textit{Direct injection} ($\rightarrow$ CVE-2024-5184) and \textit{Indirect injection} ($\rightarrow$ CVE-2025-46059). For both execution leaves, we keep $(AV,PR,UI)$ from the representative CVE and override only $AC$ using $AC_{\text{maj}}(P)$ to form the conditioned vectors. Under this shared $(AV{:}N, AC{:}L, PR{:}N, UI{:}N)$ profile, both children yield $E_{\text{Direct}}=E_{\text{Indirect}}=3.89$, and the OR aggregation gives $E(V^\star)=3.89$.

\textit{Path score and goal impact.}
The SAND aggregation (Eq.~(\ref{eq:sand})) gives the path-level exploitability for B1:
\[
E_{\text{path}}(B_1)=\min\big(E(P),\,E(V^\star)\big)=\min(3.89,\,3.89)=3.89.
\]
For $G_1$, we model a primarily integrity-driven impact with $(C,I,A)=(0.00,\,0.56,\,0.00)$, yielding an impact sub-score of $3.60$ via Eqs.~(\ref{eq:isc})--(\ref{eq:impact}). Combining this with $E_{\text{path}}$ under CVSS v3.1 (Scope: Unchanged) yields $\textit{Base}_{G_1,B_1}=7.5$ (High) via Eq.~(\ref{eq:base}).

\textit{Summary of remaining $G_1$ branches.}
We compute B2--B5 using the same OR/SAND aggregation and conditioned-execution rule. For brevity, we report their branch-level scores in Table~\ref{tab:g1_exploitability}.

\textbf{G2: Leakage of Electronic Health Record (EHR) Data.}
$G_2$ captures threats leading to unauthorized access, inference, or exfiltration of sensitive EHR data. In LLM-based systems, session memory, external tool integration, and cached inference increase the risk of both accidental and adversarial disclosure (Figure~\ref{adt_g2}). As for $G_1$, we apply our ADT--CVSS procedure, but summarize the path-level results in Table~\ref{tab:g2_exploitability}.
All $G_2$ paths share a confidentiality-only impact model with $(C,I,A)=(0.56,\,0.00,\,0.00)$ (Scope: Unchanged), yielding an impact sub-score of $3.60$. Thus, differences in final base scores are driven purely by exploitability $E_{\text{path}}$.

\textbf{G3: Disruption of Access to EHR Data.}
$G_3$ captures attacks that prevent legitimate users from accessing EHR data in a timely or reliable manner (e.g., during emergency care). We fix goal-level impact at availability-only $(C,I,A)=(0.00,\,0.00,\,0.56)$ (Scope: Unchanged) and compute $E_{\text{path}}$ using the same OR/SAND aggregation rules from Section~3. In $G_3$, multiple branches share a common precondition family (e.g., missing rate limiting, weak authentication, and unthrottled input interfaces), and branch-level differences primarily arise from the execution step. Table~\ref{tab:g3_exploitability} summarizes the resulting CVSS v3.1 scores for the main overload and denial-of-service branches.

\subsection{Risk Treatment}
\label{sec:risk_treatment}

This subsection addresses RQ3 by instantiating our ADT–CVSS framework with concrete defense sets and comparing their impact on path-level exploitability under an explicit cost model

\subsection{Risk Treatment for Prompt Injection Attack Path (G1)}
\label{sec:risk_treatment_g1}

We apply the risk-treatment procedure to the $G_1$--B1 prompt-injection path, modeled as
$X=\texttt{SAND}(\mathit{P}\rightarrow \mathit{V})$, where $\mathit{P}$ captures alternative ways to compromise the prompt channel and $\mathit{V}$ captures prompt delivery/execution (direct or indirect injection). Our objective is to reduce the path exploitability $E_{\text{path}}$ while holding the goal-level impact for $G_1$ fixed at $(C,I,A)=(0.00,0.56,0.00)$.

\paragraph{\textbf{Precondition hardening (leaf-level transforms})}
At the precondition layer, we begin from the five baseline leaves in B1-\emph{user hijacking}, \emph{user--machine hijacking}, \emph{malicious user}, \emph{stolen session/API keys}, and \emph{MiTM on the web session}. Each leaf $\ell$ is mapped to a representative CVE with a baseline CVSS vector $(AV_{\ell},AC_{\ell},PR_{\ell},UI_{\ell})$. As shown in Section 4.2, OR aggregation yields $E(\mathit{P})=3.89$ and $AC_{\mathrm{maj}}(\mathit{P})=L$. We then apply concrete controls and recompute each $E(\ell)$ using the same CVSS v3.1 weights, interpreting controls as transforms to $AV$, $AC$, $PR$, and/or $UI$.

\emph{User hijacking.}
Strong multi-factor authentication (e.g., FIDO2/WebAuthn) increases the privileges required to successfully impersonate a clinician, modeled as $PR:L\rightarrow H$, reducing $E$ from $2.84$ to $1.23$. Short-lived session tokens further increase the effective attack complexity by narrowing the exploitation window, modeled as $AC:L\rightarrow H$; the combined configuration $(PR:H,AC:H)$ yields $E=0.71$. We treat MFA as a cross-team deployment (\textit{cost level~3}) and short-lived tokens as primarily application/configuration changes (\textit{cost level~2}).

\emph{User--machine hijacking.}
Endpoint Detection and Response (EDR) and OS hardening make endpoint compromise less reliable for commodity malware, modeled as $AC:L\rightarrow H$, reducing $E$ from $3.89$ to $2.22$. Device binding and session attestation (e.g., key-bound sessions, posture checks) further restrict viable attacks and effectively require additional local capability, modeled as $PR:N\rightarrow L$ under $AC:H$, yielding $E=1.62$. We assign EDR to \textit{cost level~2} and device binding/attestation to \textit{level~3} due to enforcement and infrastructure requirements.

\emph{Malicious user.}
Role-Based Access Control (RBAC) and least-privilege policies reduce the ability of low-privilege accounts to reach high-impact clinical functions, modeled as $PR:L\rightarrow H$, lowering $E$ from $2.84$ to $1.23$. Prompt monitoring is primarily detective and does not change the base CVSS vector for this leaf; we treat it as a complementary control that affects detection/response rather than exploitability. RBAC is categorized as \textit{cost level~2}.

\emph{Stolen session/API keys.}
Scope limiting (fine-grained permissions, per-tenant keys) increases the required privilege level to achieve the same clinical effect, modeled as $PR:L\rightarrow H$, reducing $E$ from $2.07$ to $0.90$. Secret vaulting and frequent rotation reduce dwell time but do not increase the effort for a single successful exploit; we therefore do not model them as changes to $E$. Scope limiting is \textit{cost level~2}.

\emph{MiTM on the web session.}
The baseline vector $(AV:N,AC:H,PR:N,UI:N)$ yields $E=2.22$. Enforcing TLS~1.3 with certificate pinning increases the attacker capability required for practical interception (e.g., requiring compromise of a trusted anchor), modeled as $PR:N\rightarrow L$, giving $E=1.62$. Mutual TLS (mTLS) further strengthens endpoint authentication and certificate handling, modeled as $PR:L\rightarrow H$, yielding $E=0.71$. We treat TLS hardening as \textit{cost level~2} and mTLS as \textit{level~3} due to PKI lifecycle and client-certificate management.

\paragraph{\textbf{Execution hardening (guardrails on $\mathit{V}$)}}
The execution node $\mathit{V}$ represents crafting and delivering malicious prompts (directly or indirectly). In our model, if no execution-specific guardrails are deployed, $\mathit{V}$ is conditioned on the precondition family by exporting $AC_{\mathrm{maj}}(\mathit{P})$ to the execution step, yielding $\mathit{V}^{\star}$ (Section~\ref{cvss-method}). Concretely, when $AC_{\mathrm{maj}}(\mathit{P})=L$, the execution remains low-complexity and $E(\mathit{V}^{\star})=3.89$; if precondition hardening shifts $AC_{\mathrm{maj}}(\mathit{P})$ to $H$, then $E(\mathit{V}^{\star})=2.22$ even without additional execution controls. We additionally model prompt guardrails as transforms to $(AC_{\mathit{V}},UI_{\mathit{V}})$.

Canonicalization/paraphrasing and normalization (e.g., Unicode normalization, retokenization) disrupt brittle trigger patterns and are modeled as $AC_{\mathit{V}}:L\rightarrow H$, reducing $E(\mathit{V}^{\star})$ to $2.22$ (\textit{cost level~2}). Delimited input channels (structured tool inputs, fenced instruction fields) improve data/instruction separation~\cite{PPA} and are also modeled as maintaining $AC_{\mathit{V}}:H$ at \textit{cost level~1}. Sandwich-prevention rules and policy allowlists/filters restrict explicit override directives, maintaining $AC_{\mathit{V}}$ at High while adding tuning and operational overhead (\textit{cost levels~2--3}). Finally, human-in-the-loop (HITL) gating for high-impact actions increases required user interaction, modeled as $UI_{\mathit{V}}:N\rightarrow R$, which reduces $E(\mathit{V}^{\star})$ from $2.22$ to $1.62$ once $AC_{\mathit{V}}$ is already High. When HITL is deployed as an approval gate in the critical path, we treat it as \textit{cost level~4} due to staffing and sustained operational burden.

\textbf{Scenario-level outcomes and trade-offs.}
Table~\ref{tab:pi-scenarios-final} reports the baseline and four post-treatment scenarios for the $G_1$--B1 prompt-injection path.
Hardening a single precondition (S1) yields only a limited reduction because the OR aggregation over preconditions remains dominated by other network-reachable leaves, so $E(P)$ remains the limiting term.
In contrast, bundling precondition controls (S2) raises the majority precondition complexity to $AC_{\text{maj}}(P)=H$ and reduces the path score to $E_{\text{path}}=1.62$, reflecting an infrastructure-first posture.
A complementary strategy is execution-only hardening (S3): even with unchanged preconditions, guardrails that increase execution complexity reduce $E(V^\star)$ and therefore lower $E_{\text{path}}$ to 2.22, offering a pragmatic mitigation when platform changes are constrained.
Finally, adding HITL gating (S4) provides a last-resort control for high-impact clinical actions; while it reduces $E(V^\star)$, it does not improve on S2 in $E_{\text{path}}$ because the SAND composition remains bounded by $E(P)$, but it strengthens operational safety by inserting explicit human approval at the execution stage.

A useful outcome of the treatment analysis is that it makes \emph{efficiency} explicit: the SAND aggregation $E_{\text{path}}=\min(E(P),E(V^\star))$ identifies the current bottleneck and therefore the most leverageful place to invest engineering effort.
Once one stage is sufficiently hardened, additional controls on the other stage yield diminishing returns in $E_{\text{path}}$ (e.g., S2 and S4).
Rather than being a limitation, this is a practical advantage: the framework helps avoid “double-paying’’ for controls that do not further reduce exploitability, while still providing a principled basis for adding procedural defenses (e.g., HITL) for governance, assurance, or fail-safe workflows.

Overall, this case study demonstrates how the ADT--CVSS workflow supports comparative treatment analysis across heterogeneous controls: it makes explicit which base metrics are altered, how those changes propagate to $E_{\text{path}}$, and where reductions saturate due to OR/SAND aggregation.
For space, we apply the same treatment procedure to $G_2$ and $G_3$ at a summarized level.

\subsection{Risk Treatment for Orchestration Errors (G1)}
\label{sec:risk_treatment_orch}

We instantiate our treatment workflow on the $G_1$--B3 Orchestration 
Error path, modeled as $\texttt{SAND}(\mathit{P}\rightarrow 
\mathit{V})$, keeping the goal-level impact fixed at 
$(C,I,A)=(0.00,0.56,0.00)$ and quantifying how concrete controls 
reduce $E_{\text{path}}$. The preconditions $\mathit{P}$ capture 
orchestration weaknesses---lack of integrity checks, weak 
authentication, and unvalidated shared context---while the execution 
node $\mathit{V}$ captures runtime manipulations such as task 
injection, task mismanagement, memory poisoning, and 
agent-in-the-middle. We treat defenses as metric transforms on 
$(AV,AC,PR,UI)$: precondition hardening centers on signing and 
verifying orchestrator artifacts (raising $AC$ and $PR$ by requiring 
the attacker to compromise a signing key), mTLS for certificate-bound 
service identity (raising $PR$ versus token replay), and context 
segmentation plus write RBAC (raising $AC$ and/or $PR$ by enforcing 
isolation and controlled writes). Execution hardening similarly 
constrains runtime behavior via task allowlists that restrict which 
tasks the orchestrator may execute, sequence enforcement 
(DAG/state-machine checks), integrity-checked context hand-offs 
between pipeline stages (using checksums or signatures), and response 
signing for agent outputs; collectively, these controls shift the 
path from surfaces that are easily exploitable over the network 
without credentials toward interfaces that require verified identities 
and integrity checks, yielding substantial reductions in the 
representative exploitability terms. We summarize the resulting 
treatment portfolios, costs, and the corresponding 
$(E(\mathit{P}), AC_{\mathrm{maj}}(\mathit{P}), 
E(\mathit{V}^{\star}), E_{\text{path}})$ outcomes across canonical 
scenarios in Table~\ref{tab:orch-scenarios-final}.

%% file: sections/discussion.tex
\section{Discussion}
\label{sec:discussion}

It is essential to distinguish between notions such as ’attack’ and ’threat’ to understand early-stage security requirements development efforts. Compared to attacks, threats are more abstract concepts
that define the security objectives that various attacks can realize. In the early stages, threats can be
conceptually defined; however, no detailed system evidence exists to predict attack details. Threat
modeling efforts typically elicit threats individually without considering their interdependencies.
However, rigorous security risk assessment requires a more system view, in the form of a
threat path, that relates threats to each other depending on their pre- and post-conditions
and the attack actor’s objectives. Threat paths, rather than individual threats, could be more
instrumental in identifying effective and efficient countermeasures that complement each other.
Although some approaches, such as PASTA \cite {ucedavelez2015risk} or the NIST Threat Modeling Approach \cite{pulivarti2024cybersecurity},
apply attack trees to provide this system view, they address later development stages rather than
the early-stage requirement analysis phase. In this study, we propose an approach to consolidate individual threats obtained from conventional, adversarial, and conversational categories for an LLM-based system and create threat paths to achieve a specific goal. We still use the term "attack path" throughout the paper, as we use the attack tree as the main approach for identifying paths, and readers may be more familiar with this term. However, each tree node defines a threat rather than an attack. Thus, those attack paths can also be termed as threat paths. 

We utilize the CVSS framework to quantify the risk of threat paths derived from attack-defense trees. Specifically, the exploitability sub-score is used to estimate the likelihood of individual threat paths, while the impact sub-score, obtained from the root node, complements the overall risk calculation. Vulnerability databases such as the National Vulnerability Database (NVD), which are built on the CVE and CVSS frameworks, constitute one of the most comprehensive and systematically created data sources in cyber security, extensively supported by professional communities. Although these databases and scoring frameworks has originally evolved to assess vulnerabilities in deployed operational systems, we still found them instrumental in estimating risks in early-stage. In particular, CVSS has well-established exploitability sub-scores that enable us to approximate the likelihood of threat paths. Many vulnerabilities relevant to specific threat categories have been already scored by the community. We leverage the exploitability of those vulnerabilities to approximate the exploitability score of the relevant threats identified during the early stage analysis. We consider that risk quantification without relying on such a comprehensive data source would have been inherently vague.

Our results show a striking concentration of path-level CVSS base scores around 7.5 across many attack branches under $G_1$--$G_3$. At first glance, this might suggest that the framework lacks resolution, but it is in fact a direct consequence of two deliberate modeling choices. First, we focus on realistic, Internet- or network-reachable attack surfaces for LLM-powered healthcare systems, where adversaries typically enjoy network access ($AV{:}N$), face relatively low attack complexity ($AC{:}L$), and do not require prior privileges ($PR{:}N$) or user interaction ($UI{:}N$). Second, for each goal we assign a single “dominant” CIA dimension a High impact value (integrity for $G_1$, confidentiality for $G_2$, availability for $G_3$), while leaving the other dimensions at None. Under CVSS~v3.1’s discrete weights and rounding rules, many combinations of these vectors converge to a base score near 7.5, even when the internal attack logic, preconditions, and affected components differ. 

The attack–defense trees compensate for this coarse numerical clustering by exposing \emph{where} the effort lies within each path. For example, both Prompt Injection (B1) and Orchestration Errors (B3) score 7.5 under $G_1$, but the trees show that B1 is dominated by user endpoint compromise and weak prompt execution guardrails, whereas B3 concentrates risk in orchestration-layer integrity and context isolation. Similarly, under $G_2$, User Session Compromise, LLM Session Compromise, and Model Extraction all reach 7.5, yet the underlying attack families (session management vs.\ KV-cache isolation vs.\ model memorization and query controls) are structurally distinct. In this sense, the absolute base score is less important than (i) the \emph{relative} change in $E_{\text{path}}$ when we apply defenses (our $\Delta E$), and (ii) the decomposition into precondition and execution subtrees that tells an engineer where to spend effort. The quantitative layer is therefore primarily a ranking and comparison tool; identical base scores should not be interpreted as implying that risks are operationally equivalent, share the same underlying weaknesses, or admit the same mitigation strategy. 

Our risk assessment is conducted at an early stage, with no mitigations assumed, and the scenarios are selected from inherently high-impact ones; the initial findings, thus, converge to similar high values. As mitigation alternatives are included in the security requirement list, the reduced scores span a range of values, allowing us to compare them and apply more effective yet less costly mitigation plans.

Our choice of CVSS~v3.1 as the underlying scoring scheme reflects its current adoption in vulnerability management workflows and its compatibility with existing security guidance for healthcare and cloud systems. However, some LLM-specific phenomena, such as prompt flooding and resource-exhaustion via recursive reasoning, are not yet captured by official v3.1 vectors and only recently began to appear in CVE records. For one such case (prompt flooding), we derived a representative vector by aligning with emerging CVSS~v4.0 semantics, while still mapping back to the v3.1 metric set for consistency. The main limitation is that CVSS v3.1 has no native way to represent LLM-service behaviors such as cost amplification (prompt flooding), recursive agent/tool loops, or multi-tenant inference artifacts. Consequently, we must approximate these with proxies in $(AV,AC,PR,UI)$, which compresses distinct LLM attack families into similar exploitability values and reduces ranking resolution. CVSS~v4.0 introduces more granular notions of attack requirements and environment, and it is better suited to modeling multi-tenant, service-heavy architectures where LLMs reside. A natural extension of our work is therefore to re-parameterize the same attack–defense trees using full v4.0 vectors, and to compare how path rankings and treatment recommendations shift when richer metrics (e.g., safety requirements, system-specific environmental weighting) become first-class citizens.

An organizational barrier observed in the real-world security development life cycles is the presence of silos between the security, software engineering, and ML teams. Experts in these domains often lack a common understanding and language on security threats and relevant mitigations. Security engineers may be less familiar with adversarial threats and relevant mitigation techniques, while ML and software engineering experts may not comprehend the cyber kill chains that harmonize adversarial and conventional threats. Addressing these gaps requires broader interdisciplinary collaboration and security awareness programs across these teams. However, this paper provides a risk management instrument that enables these teams to consolidate and compare all potential threats and mitigations within a unified framework. Thus, the proposed framework may facilitate collaborative risk management efforts between these teams.

We also see a natural next step in using LLMs themselves to assist with threat modeling and risk analysis. In our study, the LLM is only the object under evaluation, but the same technology could be used to draft candidate attack–defense trees from natural-language system descriptions, suggest missing preconditions or defenses, or help explore large spaces of treatment scenarios by proposing concrete metric transforms and rough cost estimates. Any such use would remain advisory and require human validation, but it could substantially lower the barrier for non-expert teams to apply structured, goal-driven risk assessment to LLM-powered healthcare systems.

%% file: sections/conclusion.tex
\section{Conclusion}

We presented a goal-driven, reproducible risk-assessment framework for LLM-powered systems. Using Attack–Defense Trees across three goals (G1–G3), we factorize each path into Preconditions and Execution via SAND and aggregate with AND/OR to make threat evolution explicit. We bind CVSS v3.1 exploitability to leaves and reserve impact for goal nodes, cleanly separating “how easy” from “how bad,” which yields auditable, comparable scores. Our risk treatment recomputes path exploitability under concrete defense vector transforms and ranks options by a transparent 1–4 cost scale, enabling cost-aware mitigation choices. Although instantiated for healthcare, the workflow is plug-and-play for other LLM deployments, supports goal swaps, updates, and reruns, and supports repeatable evaluations and iterative calibration as evidence accumulates.

%% file: appendix.tex
\clearpage
\onecolumn

\begin{figure}[p]
    \centering
    \rotatebox{270}{\includegraphics[width=\textheight]{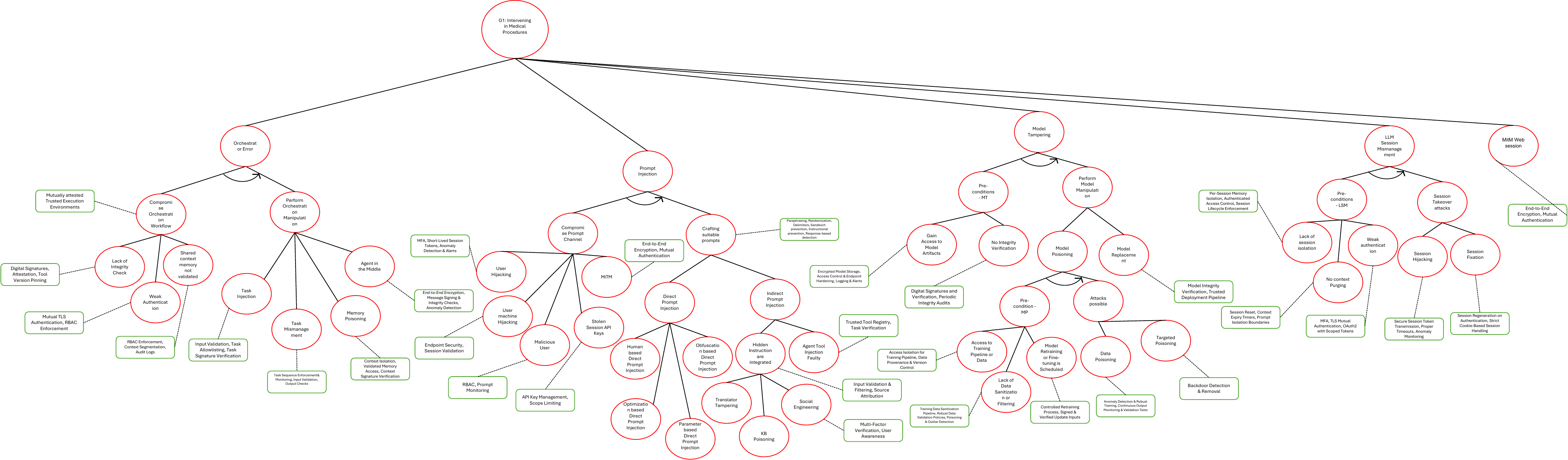}}
    \caption{Attack Tree for G1: Intervening in Medical Procedures.}
    \label{adt_g1}
\end{figure}

\begin{figure}[p]
    \centering
    \rotatebox{270}{\includegraphics[width=\textheight]{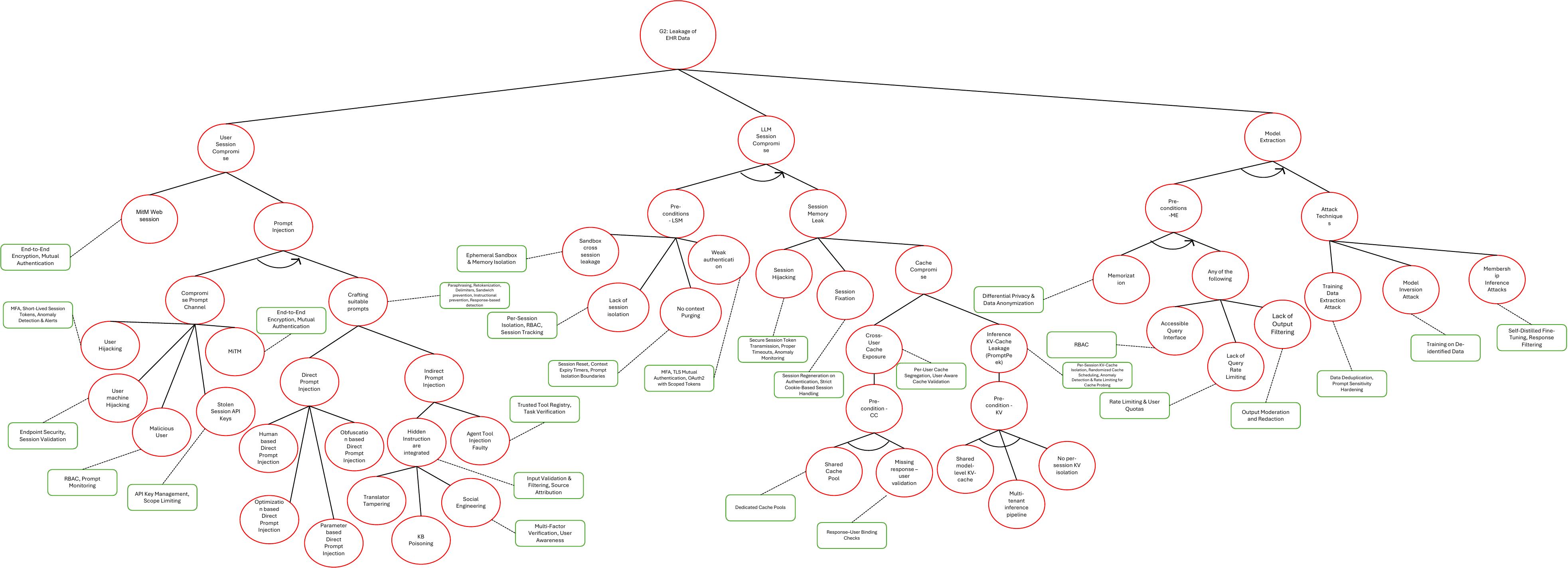}}
    \caption{Attack Tree for G2: Leakage of Electronic Health Record (EHR) Data.}
    \label{adt_g2}
\end{figure}

\begin{figure}[p]
    \centering
    \rotatebox{270}{\includegraphics[width=\textheight]{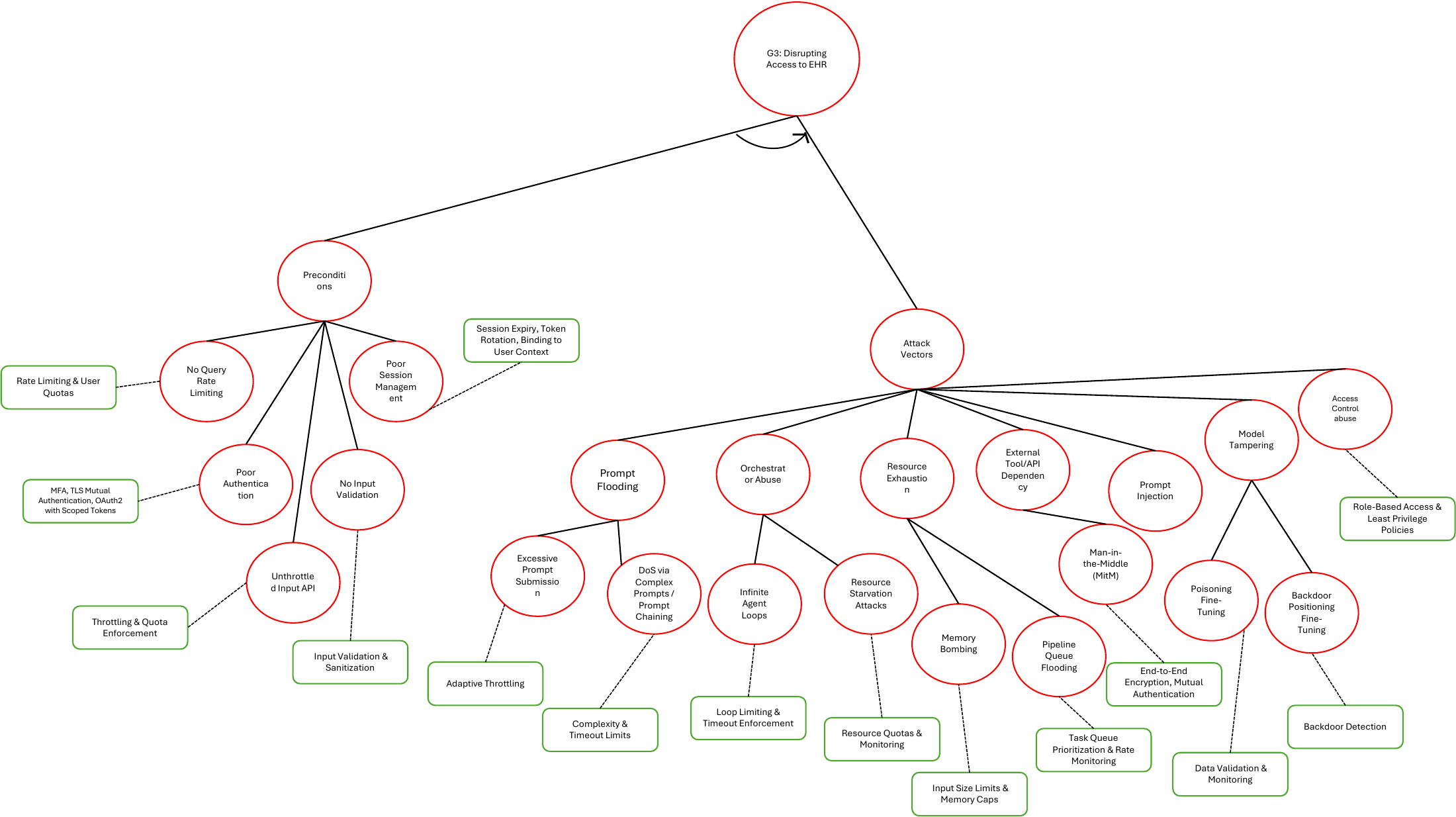}}
    \caption{Attack Tree for G3: Disruption of Access or Availability.}
    \label{adt_g3}
\end{figure}

\begin{table}[t]
\centering
\renewcommand{\arraystretch}{1.2}
\setlength{\tabcolsep}{3pt}
\begin{tabular}{|p{3.6cm}|p{3.6cm}|p{3.8cm}|c|c|c|c|}
\hline
\textbf{Branch} 
& \textbf{Precondition CVEs (examples)} 
& \textbf{Execution CVEs (examples)} 
& $E_{\text{path}}$ 
& $AC_{\text{maj}}$ 
& $(C,I,A)$ 
& \textbf{Base (S:U)} \\
\hline
B1: Prompt Injection 
& User hijacking (CVE-2025-22222); User machine hijack (CVE-2025-22457/53770/32433); Malicious user (CVE-2025-42957); Stolen API key (CVE-2024-12012); HTTPS MiTM (CVE-2024-50691)
& Direct injection (CVE-2024-5184); Indirect injection (CVE-2025-46059)
& 3.89 
& Low 
& (0.00, 0.56, 0.00) 
& 7.5 (High) \\
\hline
B2: LLM Session Mismanagement 
& Weak authentication (CVE-2025-46801); Lack of session isolation (CVE-2025-32441); No context purging (CVE-2023-4969)
& Session hijacking (CVE-2024-6587); Session fixation (CVE-2024-38513)
& 3.89 
& Low 
& (0.00, 0.56, 0.00) 
& 7.5 (High) \\
\hline
B3: Orchestration Errors 
& Lack of integrity checks (CVE-2025-59334); Weak orchestration API auth (CVE-2025-46801); Unvalidated shared context (CVE-2023-29374)
& Task injection (CVE-2024-8156); Task mismanagement (CVE-2022-29164); Memory poisoning (CVE-2024-8309); Agent-in-the-middle (CVE-2025-6159)
& 3.89 
& Low 
& (0.00, 0.56, 0.00) 
& 7.5 (High) \\
\hline
B4: Model Tampering 
& Access to model artifacts (CVE-2023-6018); Lack of integrity verification (CVE-2025-59334)
& Poisoning (CVE-2024-5185); Model replacement (CVE-2025-24357); Model extraction (CVE-2024-6868)
& 3.89 
& Low 
& (0.00, 0.56, 0.00) 
& 7.5 (High) \\
\hline
B5: Web MiTM 
& ---
& HTTPS MiTM (CVE-2024-50691)
& 2.22 
& High 
& (0.00, 0.56, 0.00) 
& 5.9 (Medium) \\
\hline
\end{tabular}
\caption{Summary of exploitability and impact for $G_1$ (Intervening in medical procedures). 
Precondition CVEs map to $P$, execution CVEs map to $V$, and $E_{\text{path}}$ is computed for $P\ \texttt{SAND}\ V$ with Scope: Unchanged.}
\label{tab:g1_exploitability}
\end{table}


\begin{table}[t]
\centering
\renewcommand{\arraystretch}{1.2}
\setlength{\tabcolsep}{3pt}
\begin{tabular}{|p{2.4cm}|p{3.8cm}|p{3.8cm}|c|c|c|c|}
\hline
\textbf{Branch} 
& \textbf{Precondition CVEs (examples)} 
& \textbf{Execution CVEs (examples)} 
& $E_{\text{path}}$ 
& $AC_{\text{maj}}$ 
& $(C,I,A)$ 
& \textbf{Base (S:U)} \\
\hline
B1: User Session Compromise
& \textit{Reused from $G_1$ (no additional precondition family in $G_2$)}
& MitM on web session (CVE-2024-50691, $E{=}2.22$, $AC_{\text{maj}}{=}$High) \textbf{OR} Prompt Injection (Direct/Indirect; CVE-2024-5184 / CVE-2025-46059, $E{=}3.89$, $AC_{\text{maj}}{=}$Low)
& 3.89
& \textit{-- (OR reuse)}
& (0.56, 0.00, 0.00)
& 7.5 (High) \\
\hline
B2: LLM Session Compromise
& Weak authentication (CVE-2025-46801, $E{=}3.89$); Lack of session isolation (CVE-2025-32441, $E{=}1.62$); No context purging (CVE-2023-4969, $E{=}1.83$); Cross-session leakage (CVE-2024-27935, $E{=}3.89$)
& \textit{Reused from $G_1$:} Session hijacking (CVE-2024-6587, $E{=}3.89$), Session fixation (CVE-2024-38513, $E{=}3.89$); \textit{Cache compromise:} Cross-user cache exposure (CVE-2023-2885, $E{=}2.84$) \textbf{OR} KV-cache leakage / PromptPeek-style (CVE-2024-27935, $E{=}3.89$)
& 3.89
& Low
& (0.56, 0.00, 0.00)
& 7.5 (High) \\
\hline
B3: Model Extraction (Training Data Leakage)
& Accessible query interface (CVE-2024-32965, $E{=}3.89$); Missing rate limiting (CVE-2024-51557, $E{=}2.84$); Insufficient output filtering (CVE-2024-3402, $E{=}2.07$)
& Extraction / inversion / membership inference (\textit{no CVE})\textsuperscript{$\dagger$}
& 3.89
& Low
& (0.56, 0.00, 0.00)
& 7.5 (High) \\
\hline
\end{tabular}
\caption{Summary of exploitability and impact for $G_2$ (Leakage of EHR data). 
All paths use confidentiality-only impact with Scope: Unchanged. B1 reuses $G_1$ exploitability values for MitM and Prompt Injection and aggregates them via OR. In B2, session hijacking and fixation reuse the $G_1$ execution CVEs, while preconditions and cache-compromise mechanisms are instantiated for $G_2$.\\
\noindent\textsuperscript{$\dagger$}We treat model memorization of EHR data as a scenario-level gating assumption (the risk applies only when memorization is present). CVSS v3.1 exploitability metrics focus on attacker reach and effort (AV, AC, PR, UI), so memorization is handled in the goal/scenario definition rather than encoded as a CVSS exploitability vector.}
\label{tab:g2_exploitability}
\end{table}


\begin{table}[t]
\centering
\renewcommand{\arraystretch}{1.2}
\setlength{\tabcolsep}{3pt}
\begin{tabular}{|p{2.6cm}|p{2cm}|p{3.8cm}|c|c|c|c|}
\hline
\textbf{Branch} 
& \textbf{Precondition CVEs} 
& \textbf{Execution CVEs} 
& $E_{\text{path}}$ 
& $AC_{\text{maj}}$ 
& $(C,I,A)$ 
& \textbf{Base (S:U)} \\
\hline
B1: Prompt flooding 
& \textsuperscript{$\dagger$}$P_{\mathrm{G3}}$ (shared) 
& CVE-2025-46569\textsuperscript{$\ddagger$} (prompt-chaining / flooding DoS, $E(V)=2.84$) 
& 2.84 
& Low 
& (0.00, 0.00, 0.56) 
& 6.5 (Medium) \\
\hline
B2: Orchestrator abuse 
& \textsuperscript{$\dagger$}$P_{\mathrm{G3}}$ (shared) 
& CVE-2025-48956 (infinite agent/tool loops, $E(V)=3.89$) 
& 3.89 
& Low 
& (0.00, 0.00, 0.56) 
& 7.5 (High) \\
\hline
B3: Resource exhaustion 
& \textsuperscript{$\dagger$}$P_{\mathrm{G3}}$ (shared) 
& CVE-2025-61771 (resource exhaustion, $E(V)=3.89$) 
& 3.89 
& Low 
& (0.00, 0.00, 0.56) 
& 7.5 (High) \\
\hline
B4: External tool / API dependency
& \textsuperscript{$\dagger$}$P_{\mathrm{G3}}$ (shared) 
& CVE-2024-2466 (external API DoS / upstream outage trigger, $E(V)=3.89$) 
& 3.89 
& Low 
& (0.00, 0.00, 0.56) 
& 7.5 (High) \\
\hline
B5: Prompt injection (availability misuse)
& \textsuperscript{$\dagger$}$P_{\mathrm{G3}}$ (shared) 
& Reused from $G_1$: CVE-2024-5184 (direct injection) / CVE-2025-46059 (indirect injection), $E(V)=3.89$ 
& 3.89 
& Low 
& (0.00, 0.00, 0.56) 
& 7.5 (High) \\
\hline
B6: Model tampering (availability impact)
& \textsuperscript{$\dagger$}$P_{\mathrm{G3}}$ (shared) 
& Reused from $G_1$ model-tampering branch (e.g., poisoning/replacement vectors), $E(V)=3.89$ 
& 3.89 
& Low 
& (0.00, 0.00, 0.56) 
& 7.5 (High) \\
\hline
B7: Access control abuse 
& \textsuperscript{$\dagger$}$P_{\mathrm{G3}}$ (shared) 
& CVE-2025-29927 (access control bypass enabling quota/resource abuse, $E(V)=3.89$) 
& 3.89 
& Low 
& (0.00, 0.00, 0.56) 
& 7.5 (High) \\
\hline
\end{tabular}
\caption{Summary of exploitability and impact for $G_3$ (Disruption of access to EHR data), with availability-only impact (Scope: Unchanged).}
\label{tab:g3_exploitability}
\vspace{0.5mm}
\footnotesize{
\noindent\textsuperscript{$\dagger$}$P_{\mathrm{G3}}$ denotes the shared precondition family from the draft: 
(i) No query rate limiting (CVE-2024-51557, $E=2.84$; reused from $G_2$), 
(ii) Poor session management (reused from $G_1$, $E=1.62$; e.g., CVE-2025-32441 vector), 
(iii) Poor authentication (reused from $G_1$, $E=3.89$; e.g., CVE-2025-46801 vector), 
(iv) No input validation (CVE-2025-8320, $E=2.84$), and 
(v) Unthrottled input APIs (CVE-2025-61770, $E=3.89$). 
This yields $E(P_{\mathrm{G3}})=3.89$ and $AC_{\text{maj}}(P_{\mathrm{G3}})=\text{Low}$.
\\
\noindent\textsuperscript{$\ddagger$}CVE-2025-46569 had no official CVSS v3.1 at the time of writing; the exploitability vector used here follows the draft’s derived mapping.
}
\end{table}

\begin{table}[t]
\centering
\renewcommand{\arraystretch}{1.15}
\setlength{\tabcolsep}{4pt}
\begin{tabular}{|c|p{5.2cm}|c|c|c|c|c|c|}
\hline
\textbf{ID} &
\textbf{Defense Set (post-treatment; after hardening)} &
$E(P)$ &
$AC_{\text{maj}}(P)$ &
$E(V^\star)$ &
$E_{\text{path}}$ &
\textbf{Final Base (S:U)} &
\textbf{Cost} \\
\hline
Baseline &
None (original path; no added controls)
& 3.89 & L & 3.89 & 3.89 & 7.5 (High) & -- \\
\hline
S1 &
\textbf{Harden one major precondition:}
User--Machine hijack $\rightarrow$ device binding / session attestation (AC:H, PR:L)
& 2.84 & L & 3.89 & 2.84 & 6.5 (Med) & 3 \\
\hline
S2 &
\textbf{Harden all preconditions:}
User Hijack: MFA + short-lived tokens (PR:H, AC:H); User--Machine: device binding (AC:H, PR:L); Malicious User: RBAC (PR:H); Stolen API Key: scope limiting (PR:H); MitM: TLS 1.3 + pinning (PR:L)
& 1.62 & H & 2.22 & 1.62 & 5.3 (Med) & 2--3 \\
\hline
S3 &
\textbf{Harden execution only (lightweight):}
Canonicalization + delimited channels + sandwich-prevention rules $(AC{:}H)$
& 3.89 & L & 2.22 & 2.22 & 5.9 (Med) & 1--3 \\
\hline
S4 &
\textbf{Harden both:}
S2 preconditions + HITL gate for high-impact actions $(UI{:}R,AC{:}H)$
& 1.62 & H & 1.62 & 1.62 & 5.3 (Med) & 2--4 \\
\hline
\end{tabular}
\caption{Prompt-injection treatment scenarios for $G_1$--B1 (Scope unchanged). Baseline reports the original path scores. Rows S1--S4 report \emph{post-treatment} values recomputed after applying the listed controls as CVSS base-metric transforms at the affected leaves and/or execution step. Impact is fixed at $(C,I,A)=(0.00,0.56,0.00)$ (Impact $=3.60$), so differences in Base are driven by $E_{\text{path}}$. Cost is reported as ordinal level(s) per Table~\ref{tab:cost_levels}.}
\label{tab:pi-scenarios-final}
\end{table}


\begin{table}[t]
\centering
\renewcommand{\arraystretch}{1.15}
\setlength{\tabcolsep}{4pt}
\begin{tabular}{|c|p{5.6cm}|c|c|c|c|c|c|}
\hline
\textbf{ID} &
\textbf{Defense Set (post-treatment; after hardening)} &
$E(P)$ &
$AC_{\text{maj}}(P)$ &
$E(V^\star)$ &
$E_{\text{path}}$ &
\textbf{Final Base (S:U)} &
\textbf{Cost} \\
\hline
Baseline &
None (original path; no added controls)
& 3.89 & L & 3.89 & 3.89 & 7.5 (High) & -- \\
\hline
S1 &
\textbf{Harden major preconditions only:}
mTLS on orchestrator APIs $(PR{:}N\!\rightarrow\!H)$; context segmentation + write-RBAC $(AC{:}L\!\rightarrow\!H,\ PR{:}N\!\rightarrow\!L)$
& 2.84 & L & 3.89 & 2.84 & 6.5 (Med) & 2--4 \\
\hline
S2 &
\textbf{Harden all preconditions:}
sign \& verify orchestration artifacts $(AC{:}L\!\rightarrow\!H,\ PR{:}N\!\rightarrow\!L)$; mTLS on APIs $(PR{:}N\!\rightarrow\!H)$; context segmentation + write-RBAC $(AC{:}L\!\rightarrow\!H,\ PR{:}N\!\rightarrow\!L)$
& 1.62 & H & 2.22 & 1.62 & 5.3 (Med) & 3--4 \\
\hline
S3 &
\textbf{Harden execution only:}
task allowlisting + task/plan signing; task-sequence enforcement; context integrity checks; agent-response signing/verification
& 3.89 & L & 1.62 & 1.62 & 5.3 (Med) & 2--4 \\
\hline
S4 &
\textbf{Harden both:}
S2 preconditions + S3 execution controls
& 1.62 & H & 1.62 & 1.62 & 5.3 (Med) & 3--4 \\
\hline
\end{tabular}
\caption{Orchestration-error treatment scenarios for $G_1$--B3 (Scope unchanged). Baseline reports the original path scores. Rows S1--S4 report \emph{post-treatment} values recomputed after applying the listed controls as CVSS base-metric transforms at the affected leaves and/or execution step. Impact is fixed at $(C,I,A)=(0.00,0.56,0.00)$ (Impact $=3.60$), so differences in Base are driven by $E_{\text{path}}$. Cost is reported as ordinal level(s) per Table~\ref{tab:cost_levels}.}
\label{tab:orch-scenarios-final}
\end{table}